\documentclass[preprint,aps,prd,tightenlines,showpacs,nofootinbib,
superscriptaddress]
              {revtex4}%
\usepackage{graphicx}
\usepackage{amsmath}
\usepackage{bm}
\usepackage{relsize}
\RequirePackage{xspace}
\newcommand{\babar}{\mbox{\slshape B\kern-0.1em{\smaller A}\kern-0.1em %
B\kern-0.1em{\smaller A\kern-0.2em R}}\xspace}

\begin{document}

\preprint{\begin{tabular}{l}
          ANL-HEP-PR-05-19\end{tabular}}
\title{Comparison of the color-evaporation model 
	and the NRQCD factorization approach in charmonium production}


\author{Geoffrey T.~Bodwin}
\affiliation{
High Energy Physics Division, 
Argonne National Laboratory, 
9700 South Cass Avenue, Argonne, Illinois 60439}

\author{Eric Braaten}
\affiliation{
Physics Department, Ohio State University, Columbus, Ohio 43210}

\author{Jungil Lee}
\affiliation{
Department of Physics, 
Korea University, 
Seoul 136-701, Korea}


\date{\today}
\begin{abstract}
We compare the color-evaporation model (CEM) and nonrelativistic QCD
(NRQCD) factorization predictions for inclusive quarkonium production. 
Using the NRQCD factorization formulas for
quarkonium production and for perturbative $Q\bar Q$ production, we deduce
relationships that are implied by the CEM between
the nonperturbative NRQCD matrix elements that appear in the
factorization formula for quarkonium production. These relationships are
at odds with the phenomenological values of the matrix elements that
have been extracted from the Tevatron data for charmonium production at 
large transverse momentum. A direct comparison of the CEM and NRQCD
factorization predictions with the CDF charmonium production data
reveals that the CEM fits to the data are generally unsatisfactory,
while the NRQCD factorization fits are generally compatible with the
data. The inclusion of $k_T$ smearing improves the CEM fits
substantially, but significant incompatibilities remain. The NRQCD
factorization fits to the $\chi_c$ data indicate that multiple gluon
radiation is an essential ingredient in obtaining the correct shape of
the cross section as a function of $p_T$.
\end{abstract}



\maketitle


\section{Introduction}

In general, the production rates for specific hadronic states in 
high-energy processes are difficult to understand from first principles 
because they involve nonperturbative aspects of QCD in an essential 
way.  The inclusive production rates for heavy-quarkonium states have two 
aspects that facilitate their understanding from first principles.  First, 
the mass $m_Q$ of the heavy quark and the antiquark that are the 
constituents of the quarkonium are large compared to the scale
$\Lambda_{\rm QCD}$ that is associated with the nonperturbative aspects
of QCD.  Second, the inclusive nature of the quarkonium production
process may make it less sensitive, at transverse momenta $p_T$ that are
much greater than $\Lambda_{\rm QCD}$, to nonperturbative effects that
are associated with the formation of color-singlet
hadrons from the colored partons.

Two models for the inclusive production of heavy quarkonium were
introduced in the 1970's:  the color-singlet model (CSM) and the
color-evaporation model (CEM).  These models make very different
assumptions about the r\^oles played by the colors and spins of the
heavy quark and antiquark in the production process.  In the CSM, it is
assumed that a specific quarkonium state can be formed only if the $Q
\bar Q$ pair is created in a color-singlet state with the 
same angular-momentum 
quantum numbers as the quarkonium. In the CEM, the probability
of forming a specific quarkonium state is assumed to be independent of
the color of the $Q \bar Q$ pair.  In some versions of the CEM, the
probability of forming a specific quarkonium state is also assumed to be
independent of the spin of the $Q\bar Q$ pair. In spite of the very
different assumptions that are made in the CSM and CEM, both models
enjoyed considerable phenomenological success through the 1980's and
into the 1990's.

In 1995, a new approach for describing inclusive heavy-quarkonium
production based on first principles was developed:  the 
nonrelativistic QCD (NRQCD)
factorization approach \cite{Bodwin:1994jh}.  It makes use of an
effective field theory called NRQCD \cite{Caswell:1985ui,Thacker:1990bm} 
to exploit the large mass of
the heavy quark, and it further assumes that, owing to the inclusive
nature of the production process, traditional factorization methods can
be exploited to establish a factorization formula. The NRQCD
factorization approach incorporates aspects of both the CSM and CEM and
can be regarded as a unification of these two models within a consistent
theoretical framework.  It can be summarized by the NRQCD factorization
formula, which separates short-distance, perturbative effects involving
momenta of order $m_Q$ from long-distance, nonperturbative effects.

The nonperturbative factors in the NRQCD factorization formula are NRQCD
matrix elements. NRQCD can be used to give rough predictions of
the relative sizes of the matrix elements. These predictions
are based on the leading power behavior of the matrix elements as a
function of the typical heavy-quark velocity $v$ in the quarkonium rest
frame. In order to obtain an estimate, one makes the assumption that,
for each matrix element, the coefficient of the leading power of $v$ is
of order unity. This last assumption may not be reliable. However, one
can ignore this $v$-scaling information and treat the NRQCD
factorization formula as a general phenomenological framework for
analyzing inclusive heavy-quarkonium production. Any model that can be
described in terms of QCD processes at short distances, including the
CSM and the CEM, can be formulated in terms of assumptions about the
matrix elements in the NRQCD factorization formula.

In this paper, we derive relationships between the NRQCD 
nonperturbative factors that follow from the model assumptions of the 
CEM. We find that these relationships are often poorly satisfied by 
phenomenological values of the NRQCD matrix elements. Furthermore, the 
relationships sometimes violate the $v$-scaling rules of NRQCD. 
We conclude that the CEM and NRQCD provide very different 
pictures of the evolution of a heavy quark-antiquark pair into a 
quarkonium.

Using existing data, one can exclude the CSM as a quantitative model of 
heavy-quarkonium production.  In 1995, the CDF collaboration measured the
cross sections for the prompt production of $J/\psi$ and $\psi (2S)$ in
$p \bar p$ collisions at a center-of-mass energy of 1.8~TeV. It
discovered that the cross sections are more than an order-of-magnitude
larger than those predicted by the CSM.  This dramatic discrepancy
eliminated the CSM as a viable model for inclusive heavy quarkonium
production. (A detailed discussion of these comparisons between the CSM
and the experimental data and references to the published experimental
results and theoretical results can be found in
Ref.~\cite{Brambilla:2004wf}.)

The CEM can also be ruled out on the basis of the simple qualitative
prediction that the ratio of the inclusive production rates for any
two quarkonium states should be independent of the process. The most
dramatic violation of this prediction that has been observed is in the
fraction of $J/\psi$'s that come from decays of the $P$-wave charmonium
states $\chi_{c1}$ and $\chi_{c2}$.  This fraction is measured to be
$0.11 \pm 0.02$ in $B$ decays and $0.297 \pm 0.017_{\rm stat} \pm
0.057_{\rm syst}$ for prompt production at the Fermilab Tevatron. 
The version of the CEM in which the probability for the formation of a
quarkonium is assumed to be independent of the spin state of the $Q \bar
Q$ pair can be ruled out on the basis of several other simple
qualitative predictions.  One such prediction is that the inclusive
production rate of a quarkonium state should be independent of its spin
state, so that it should always be produced unpolarized. This prediction
is contradicted by the observation of nonzero polarization of $J/\psi$'s
in $e^+ e^-$ annihilation at the $B$ factories and by the observation of
nonzero polarization of the bottomonium states $\Upsilon (2S)$ and
$\Upsilon (3S)$ in a fixed-target experiment.  A further prediction of
this version of the CEM is that the production rates for the
$P$-wave charmonium state $\chi_{cj}$ should be proportional to $2j+1$,
and, hence, that the ratio of the inclusive cross sections for
$\chi_{c1}$ and $\chi_{c2}$ production should be 3/5. For prompt
production at the Tevatron, this ratio has been measured to be $1.04 \pm
0.29_{\rm stat} \pm 0.12_{\rm syst}$. (A detailed discussion of these
comparisons between the CEM and the experimental data and references to
the published experimental results and theoretical results can be found
in Ref.~\cite{Brambilla:2004wf}.)

Since the CEM is only a model, it can be salvaged simply by declaring it
to have a limited domain of applicability.  The failure of the
predictions for polarization can be avoided by declaring the model to
apply only to cross sections that are summed over the spin states of the
quarkonium. In the case of the predictions that ratios of quarkonium
cross sections should be the same for all processes, the most dramatic
failures can be avoided by declaring the model to apply only when the
total hadronic energy is sufficiently large.  This condition can be used
to exclude applications to $B$ decays and to $e^+ e^-$ annihilation at
$\sqrt {s} = $10.6 GeV.  In this case, the CEM reduces essentially to a
model for inclusive production of quarkonium, without regard to spin, in
high-energy fixed-target experiments, $p \bar p$ collisions at the
Tevatron, and $pp$ collisions at the LHC.

It may be possible to exclude even this limited version of the 
CEM, given sufficiently accurate experimental 
information.  In this paper, we present a quantitative comparison of the 
predictions of the CEM and the NRQCD factorization approach for 
transverse-momentum distributions of heavy quarkonia at the Tevatron.  
We restrict 
our attention to the region of transverse momentum comparable to or larger 
than the quarkonium mass, where the effects of multiple soft-gluon emission 
are not so important.  We find that the NRQCD factorization approach 
gives a significantly better fit to the available data.

The remainder of this paper is organized as follows. In
Sec.~\ref{sec:NRQCDfact}, we describe the NRQCD factorization formula for
quarkonium production and the $v$-scaling rules for the NRQCD matrix
elements. In Sec.~\ref{sec:CEM}, we describe the CEM, use the NRQCD
factorization formula to derive  expressions for the ratios of NRQCD
matrix elements that are implied by the CEM assumptions, and deduce
the $v$-scaling rules that follow from these ratios.
Sec.~\ref{sec:s-wave} contains a comparison of the
ratios implied by the CEM with the ratios of 
phenomenological matrix elements that have been extracted from the
Tevatron data for $J/\psi$ and $\psi(2S)$ production.
Sec.~\ref{sec:s-wave} also contains a direct comparison of fits of the
CEM  predictions and the NRQCD factorization predictions to the Tevatron data.
Sec.~\ref{sec:p-wave} contains similar comparisons for $\chi_c$ 
production. Finally, we present our conclusions in
Sec.~\ref{sec:conclusions}.

\section{NRQCD Factorization Formula}
\label{sec:NRQCDfact}

The NRQCD factorization formula for the inclusive cross section for 
production of a specific heavy-quarkonium state $H$ is
\begin{equation}
\sigma[A B \to H + X]
=\sum_n c^{AB}_n(\Lambda) \langle {\cal O}_n^H(\Lambda) \rangle.
\label{prod-nrqcd}
\end{equation}
Here, $A$ and $B$ are light hadrons, photons, or leptons, and $\Lambda$
is the ultraviolet cutoff of the effective theory. The $c_n^{AB}$ are
short-distance coefficients that can be calculated in perturbation
theory by matching amplitudes in NRQCD with those in full QCD. The
matrix elements $\langle {\cal O}_n^H \rangle$ are vacuum-expectation
values of four-fermion operators in NRQCD, evaluated in the rest frame
of the quarkonium. These matrix elements contain all of the
nonperturbative physics of the evolution of a $Q\bar Q$ pair into a
quarkonium state. The operators have the form
\begin{equation}
{\cal O}_n^H =
\chi^\dagger \kappa_n \psi {\cal P}^H(\Lambda) \psi^\dagger \kappa_n' \chi,
\label{H-ops}
\end{equation}
where $\psi$ is the two-component (Pauli) spinor that annihilates a
heavy quark, $\chi$ is the two-component spinor that creates a heavy
antiquark, 
and ${\cal P}^H$ is a projector onto states that in the 
asymptotic future contain the quarkonium $H$ plus light partons $X$
whose energies and momenta lie below the cutoff $\Lambda$ of the
effective field theory: 
\begin{equation}
{\cal P}^H(\Lambda) =
\sum_X | H+X, t \to \infty \rangle \langle H+X, t \to \infty|.
\label{P-ops}
\end{equation}
The factors $\kappa_n$ and $\kappa_n'$ in Eq.~(\ref{H-ops}) are
direct products of a color matrix (either the unit matrix or the matrix
$T^a$ with octet index $a$), a spin matrix (either the unit matrix or
the matrix $\sigma^i$ with triplet index $i$), and a polynomial in the
QCD covariant derivative $\bm{D}=\bm{\nabla}+ig\bm{A}$ and the QCD field
strengths. The NRQCD factorization formula in Eq.~(\ref{prod-nrqcd})
was proposed in Ref.~\cite{Bodwin:1994jh}.  Some hard-scattering
factorization formulas, such as those for deep-inelastic scattering,
Drell-Yan lepton-pair production, and $e^+e^-$ annihilation into
hadrons, have been proven to hold to all orders in the strong
coupling $\alpha_s$. The derivation of the NRQCD factorization formula
is not at this level of rigor.\footnote{A recent study of certain
two-loop contributions to quarkonium production \cite{Nayak:2005rw} has
revealed that, if factorization is to hold, then the color-octet NRQCD
production matrix elements must be modified from the form given in
Eq.~(\ref{H-ops}) by the inclusion of light-like eikonal lines that run
from each of the $Q\bar Q$ bilinears to the far future. It is not known
if this modification preserves the factorized form in higher orders.}
However, in this regard it is no different from the factorization
formula for semi-inclusive production of a hadron in hadron-hadron
collisions. The existing all-orders proofs of factorization formulas
require that the observed scattered particle be produced at a large
transverse momentum compared with the QCD scale $\Lambda_{\rm QCD}$ and
that the cross section be sufficiently inclusive. ``Sufficiently
inclusive'' means that the variables in which the cross section is
differential cannot assume values that restrict final-state parton
momenta in the parton-level cross section to be within order
$\Lambda_{\rm QCD}$ of soft or collinear singularities.

The matrix elements in Eq.~(\ref{prod-nrqcd}) fall into a hierarchy
according to their scaling with the velocity $v$ of the heavy quark (or
antiquark) in the quarkonium rest frame. $v^2\approx 0.3$ for
charmonium, and $v^2\approx 0.1$ for bottomonium. In practice, the
summation over these matrix elements is usually truncated at a low order
in $v$. The NRQCD factorization formalism has enjoyed a good deal of
phenomenological success in describing inclusive quarkonium production
at hadron, $ep$, and $e^+e^-$ colliders and in fixed-target
experiments.\footnote{See Ref.~\cite{Brambilla:2004wf} for a recent summary
of the phenomenology of quarkonium production.}

A standard set of the NRQCD operators ${\cal O}_n^H$ that appear
naturally in cross sections that are summed over the spin states of the
quarkonium was introduced in Ref.~\cite{Bodwin:1994jh}. They are
denoted by ${\cal O}_1^H({}^{2s+1}L_j)$ and ${\cal
O}_8^H({}^{2s+1}L_j)$, where the subscript indicates the color state of
the $Q \bar Q$ pair ($1$ for singlet and 8 for octet), and the argument
indicates the angular-momentum state of the $Q \bar Q$ pair ($s$ is the
total spin quantum number, $L =S, P, \ldots$ indicates the
orbital-angular-momentum quantum number, and $j$ is the
total-angular-momentum quantum number). There are implied sums over the
spin states of the quarkonium $H$. 

The velocity-scaling rules of NRQCD imply an intricate pattern of
suppression factors for the NRQCD matrix elements $\langle {\cal O}_n^H
\rangle$.  The suppression factors, up to order $v^4$ for $S$-wave and
$P$-wave $Q \bar Q$ channels, are given in Table~\ref{tab:NRQCD} for the
representative $S$-wave multiplet that consists of the charmonium states
$\eta_c$ and $J/\psi$ and for the representative $P$-wave charmonium
multiplet that consists of the charmonium states $h_c$, $\chi_{c0}$,
$\chi_{c1}$, and $\chi_{c2}$.

\begin{table}
\caption{ Velocity-suppression factors for NRQCD matrix elements in
$S$-wave and $P$-wave $Q \bar Q$ channels in NRQCD and in the CEM. The 1
or 8 indicates the color channel and ${}^{2s+1}L_j$ indicates the
angular-momentum channel. For NRQCD, the $v$-suppression factors up
to order $v^4$ are given for representative $S$-wave and $P$-wave
multiplets.  For the CEM, the orders of the $v$-suppression factors
are independent of the quarkonium state $H$, as is described
in Sec.~\ref{sec:CEM}. }
\label{tab:NRQCD}
\begin{ruledtabular}
\begin{tabular}{c|cccc|cccccccc}
& 
$\;1,{}^1S_0\;$ & $\;1,{}^3S_1\;$ & $\;8,{}^1S_0\;$ & $\;8,{}^3S_1\;$ &
$\;1,{}^1P_1\;$ & $\;1,{}^3P_0\;$ & $\;1,{}^3P_1\;$ & $\;1,{}^3P_2\;$ &
$\;8,{}^1P_1\;$ & $\;8,{}^3P_0\;$ & $\;8,{}^3P_1\;$ & $\;8,{}^3P_2\;$ \\
\hline\hline
\multicolumn{13}{c}{NRQCD Factorization}\\
\hline
$\eta_c$    & 
1 &   & $v^4$ & $v^3$ &   &   &   &   & $v^4$ &       &       &      \\
$J/\psi$    &
  & 1 & $v^3$ & $v^4$ &   &   &   &   &       & $v^4$ & $v^4$ & $v^4$ \\
\hline
$h_c$       &
  &   & $v^2$ &      & $v^2$ &      &      &      &   &   &   &      \\
$\chi_{c0}$ &
  &   &      & $v^2$ &      & $v^2$ &      &      &   &   &   &      \\
$\chi_{c1}$ &
  &   &      & $v^2$ &      &      & $v^2$ &      &   &   &   &      \\
$\chi_{c2}$ &
  &   &      & $v^2$ &      &      &      & $v^2$ &   &   &   &      \\
\hline\hline
\multicolumn{13}{c}{Color-Evaporation Model}\\
\hline
$H$    & 
1 & 1 & 1 & 1 & $v^2$ & $v^2$ & $v^2$ & $v^2$ & $v^2$ & $v^2$ & $v^2$ & $v^2$ 
\end{tabular}
\end{ruledtabular}
\end{table}

\section{Color-Evaporation Model}
\label{sec:CEM}

The color-evaporation model (CEM) was first proposed in 1977
\cite{Fritzsch:1977ay,Halzen:1977rs,Gluck:1977zm,Barger:1979js}.
In the CEM, the cross section for production of a quarkonium state $H$ is some
fraction $F_H$ of the cross section for producing $Q \bar Q$ pairs with
invariant mass below the $M \bar M$ threshold, where $M$ is the lowest-mass 
meson containing the heavy quark $Q$. The
fractions $F_H$ are assumed to be universal so that, once they are
determined by data, they can be used to predict the cross sections in
other processes and in other kinematic regions.  The cross section for
producing $Q \bar Q$ pairs that is used in the CEM has an upper limit on
the $Q \bar Q$ pair mass, but no constraints on the color or spin of the
final state.  The $Q \bar Q$ pair is assumed to neutralize its color by
interaction with the collision-induced color field, that is, by ``color
evaporation.''  In some versions of the CEM \cite{Amundson:1996qr}, the
color-neutralization process is also assumed to randomize the spins of
the $Q$ and the $\bar Q$. The CEM parameter $F_H$ is the probability
that a $Q \bar Q$ pair with invariant mass less than $2 m_M$, where
$m_M$ is the mass of the meson $M$, will bind to form the quarkonium
$H$. That probability is assumed to be 0 if the $Q \bar Q$ pair has
invariant mass greater than $2 m_M$.

In the CEM, the production cross section for the quarkonium state $H$ in
the collisions of light hadrons, photons, or leptons $A$ and $B$ is
\begin{equation}
\sigma_{\rm CEM}[A B \to H + X]  =  F_H
\int_{4m^2}^{4m_M^2} dm^2_{Q\bar Q}\, \frac{d\sigma}{dm^2_{Q\bar Q}}
[AB\to Q\bar Q+X],
\label{prod-cem}
\end{equation} 
where $m_{Q\bar Q}$ is the invariant mass of the $Q\bar Q$ pair, $m$ is
the heavy-quark mass, and $d\sigma/dm^2_{Q\bar Q}$ on the right side is
the inclusive differential cross section for a $Q\bar Q$ pair to be
produced in a collision of $A$ and $B$.  There is an implied sum over the
colors and spins of the final-state $Q\bar Q$ pair. This is where the
central model assumptions of color evaporation and spin randomization
manifest themselves.

If $A$ and/or $B$ are hadrons or photons,
the cross section for $AB \to Q \bar Q + X$
can be expressed as convolutions of parton distributions for 
$A$ and/or $B$ and a parton cross section.
At leading order in $\alpha_s$, the parton process 
$i j \to Q \bar Q$ creates a $Q \bar Q$ pair 
with zero transverse momentum,
and the differential cross section $d \sigma/dp_T^2$
is proportional to $\delta(p_T^2)$.
At next-to-leading order in $\alpha_s$ (NLO), there are  
parton processes $i j \to Q \bar Q+k$ that create a $Q \bar Q$ pair 
with nonzero $p_T$.  The complete NLO differential cross section 
is a distribution that includes singular terms proportional to 
$\delta(p_T^2)$ and $1/p_T^2$, 
but whose integral over $p_T^2$ is well behaved.
Some kind of smearing over $p_T$ is necessary to obtain a smooth 
$p_T$ distribution that can be compared with experiment.
The physical origin of the smearing is multiple gluon emission
from the initial- and final-state partons. A rigorous treatment of the
effects of multiple gluon emission requires the resummation of
logarithmic corrections to all orders in $\alpha_s$ 
\cite{Collins:1981uk,Collins:1981va,Berger:2004cc}. A simple
phenomenological model for the effects of multiple gluon emission is
$k_T$ smearing, in which the colliding partons are given Gaussian
distributions in the intrinsic transverse momentum with a width $\langle
k_T^2 \rangle$ that is treated as a phenomenological parameter.

Complete NLO calculations of quarkonium production in hadronic
collisions using the CEM have been carried out in
Refs.~\cite{Gavai:1994in,Schuler:1996ku}, using the exclusive $Q \bar Q$
production code of Ref.~\cite{Mangano:kq} to obtain the $Q \bar Q$ pair
distributions.  There are also calculations in the CEM beyond LO that
use only a subset of the NLO diagrams~\cite{Amundson:1996qr} and
calculations that describe the soft color interaction within the
framework of a Monte Carlo event generator~\cite{Edin:1997zb}.
Calculations beyond LO in the CEM have also been carried out for $\gamma
p$, $\gamma\gamma$, and neutrino-nucleon collisions and for $Z$ decays
\cite{Eboli:1998xx,Eboli:2003fr,Eboli:2003qg,Eboli:2001hc,Gregores:1996ek}.

We now proceed to elucidate the relationship between the CEM and
the NRQCD factorization formula. According to the NRQCD factorization
formalism, the differential cross section for the process $AB\to Q\bar
Q+X$ is given by
\begin{equation}
\frac{d\sigma}{dm_{Q\bar Q}^2}[AB\to Q\bar Q+X]= 
\sum_n c^{AB}_n
\sum_{\rm spins} \sum_{\rm colors} \int\frac{d^3k}{(2\pi)^3}\,
\langle {\cal O}_n^{Q(+\bm{k}) \bar Q(-\bm{k})}
\rangle\, 
\delta[m_{Q\bar Q}^2-4(\bm{k}^2+m^2)],
\label{Q-barQ-prod}
\end{equation}
where the $c^{AB}_n$ are the same short-distance coefficients that
appear in Eq.~(\ref{prod-nrqcd}). We have suppressed the dependence of
the coefficients and the operators on the ultraviolet cutoff $\Lambda$
of the effective field theory. The operator ${\cal O}_n^{Q(+\bm{k}) \bar
Q(-\bm{k})}$ is analogous to the operator in Eq.~(\ref{H-ops}), except
that the quarkonium state $H$ is replaced by a perturbative state that
consists of a $Q$ and a $\bar Q$ with momenta $\pm {\bm k}$ and with
definite spin and color indices that have been suppressed. Those
suppressed indices are summed over in Eq.~(\ref{Q-barQ-prod}). It is
convenient to define a $Q \bar Q$ operator that includes the sum over
the colors and spins and an average over the directions of the momenta
$\pm \bm{k}$ of the $Q$ and $\bar Q$ :
\begin{eqnarray}
{\cal O}_n^{Q\bar Q} (k) &=&
\chi^\dagger\kappa_n\psi 
\left( \int \frac{d \Omega_k}{4 \pi} 
	\sum_{\rm spins} \sum_{\rm colors}
{\cal P}^{Q(+\bm{k}) \bar Q(-\bm{k})}
\right)
\psi^\dagger \kappa_n' \chi,
\label{Q-barQ-ops}
\end{eqnarray}
where $d\Omega_k$ is the element of angular integration of $\bm{k}$
and $k=|\bm{k}|$. 

Comparing Eqs.~(\ref{prod-nrqcd}) and (\ref{prod-cem}) and making use of 
Eq.~(\ref{Q-barQ-prod}), we see that the CEM implies that
\begin{equation}
\sum_n c^{AB}_n \langle {\cal O}_n^H \rangle
= F_H \sum_n c^{AB}_n
\frac{1}{2\pi^2}
\int_0^{k_{\rm max}}k^2 dk\,
\langle {\cal O}_n^{Q\bar Q}(k) \rangle,
\label{nrqcd-cem}
\end{equation}
where $k_{\rm max}=\sqrt{m_M^2-m^2}$. Eq.~(\ref{nrqcd-cem}) is the
central relation that connects the NRQCD factorization approach with the
CEM. If NRQCD factorization can be established to all orders in
perturbation theory, then Eq.~(\ref{nrqcd-cem}), which is a statement
about the perturbative production of $Q\bar Q$ states, must hold
rigorously. This is to be contrasted with the NRQCD factorization
approach itself, in which nonperturbative effects could conceivably
spoil the factorization, even if the factorization formula can be
established to all orders in perturbation theory. Note, however, that
there is an implicit assumption in Eq.~(\ref{nrqcd-cem}) that the sum
over $n$ converges. The sum need not converge rapidly in order to
establish the relationship between the NRQCD factorization approach and
the CEM, but it must converge. The expansion parameter in the sum is
$k_{\rm max}^2/m^2\sim v^2$. Hence, if the sum is to converge, we must
have $k_{\rm max}^2/m^2 \lesssim 1$. In the charmonium system, $M$ is
the $D$ meson, with mass 1.86~GeV. Taking $m=1.5$~GeV, we find that
$k_{\rm max}^2/m^2 \approx 0.54$. In the bottomonium system, $M$ is the
$B$ meson, with mass 5.28~GeV. Taking $m=4.7$~GeV, we find that $k_{\rm
max}^2/m^2 \approx 0.26$. Thus the assumption $k_{\rm max}^2/m^2
\lesssim 1$ is satisfied for the bottomonium system, but is only
marginally satisfied for the charmonium system. There is also an implicit
assumption that the energy of the produced quarkonium is not near the
kinematic limit for the process. Near the kinematic limit, the $v$
expansion of NRQCD breaks down, and, so, one needs to carry out a
resummation of classes of NRQCD matrix elements in order to
maintain the accuracy of the calculation \cite{Beneke:1997qw}. Near the
kinematic limit, the short-distance coefficients $c_n^{AB}$ contain
large logarithms that must be resummed as well \cite{Fleming:2003gt}.

Now we wish to establish that the equality in Eq.~(\ref{nrqcd-cem})
must hold independently for each term in the sum over $n$. If we
truncate the series at a finite number of terms, as is often done in the
phenomenology of quarkonium production, then the short-distance
coefficients $c^{AB}_n$ generally vary independently of each
other as the kinematic variables of the incoming and outgoing particles
vary. However, in some processes, for example, in quarkonium production
as a function of $p_T$ at the Tevatron, some of the short-distance
coefficients show identical or nearly identical behavior as a function
of the kinematic variable(s). Then one can establish the equality only
of linear combinations of matrix elements. However, one can make a
stronger assumption that the CEM must hold, not just for a particular
quarkonium production process, but for all possible quarkonium
production processes. Then, since Eq.~(\ref{nrqcd-cem}) must hold for an
arbitrarily large number of processes that have different short-distance
coefficients, the equality in Eq.~(\ref{nrqcd-cem}) must hold
independently for each term in the sum over $n$. Under this assumption,
the CEM predicts that the NRQCD production matrix elements are given by
\begin{equation}
\langle {\cal O}_n^H \rangle =
\frac{1}{2\pi^2}F_H
\int_0^{k_{\rm max}}k^2 dk\,
\langle {\cal O}_n^{Q\bar Q}(k) \rangle.
\label{nrqcd-cem-me}
\end{equation}

The matrix elements on the right side of                   
Eq.~(\ref{nrqcd-cem-me}) can be computed in perturbation theory.
Their dependences on $\bm{k}$ are governed by the powers of the
covariant derivative $\bm{D}$ that appear in the factors $\kappa_n$ and
$\kappa_n'$ of the operators. For the matrix element of leading order in
$v$ that corresponds to an operator of orbital-angular-momentum quantum
number $l$, the matrix element is proportional to $k^{2l}$. Hence, the
integral on the right side of Eq.~(\ref{nrqcd-cem-me}) is proportional
to $k_{\rm max}^{2l+3}/(2l+3)$. Since $k_{\rm max}$ scales as $v$, the
matrix element is suppressed as $v^{2l}$ compared to the matrix element
of an $S$-wave operator. Thus, the CEM implies a velocity-suppression
pattern that is independent of the quarkonium state and depends only on
the orbital-angular-momentum quantum number of the $Q\bar Q$ pair. The
suppression pattern for $S$-wave and $P$-wave matrix elements is shown
in Table~\ref{tab:NRQCD}.  It should be contrasted with the intricate
suppression pattern implied by NRQCD. We note that the relation
(\ref{nrqcd-cem-me}) partially satisfies the constraints that are
imposed by the velocity-scaling rules of NRQCD, in that there is an
additional power of $v^2$ for each unit of orbital angular momentum in
the operator. However, powers of $v$ that do not arise from the
orbital angular momentum are not reproduced in the relation
(\ref{nrqcd-cem-me}). This fact has been noted previously by Beneke
\cite{Beneke:1997av}.

Now let us state explicitly the relationships between the NRQCD matrix
elements that are implied by the CEM relation (\ref{nrqcd-cem-me}). In
the operator ${\cal O}_n$, let the subscript $n$ represent the
angular-momentum quantum numbers ($s$, $l$, and $j$) and the color state
(singlet or octet). The standard forms of the operators are given in
Ref.~\cite{Bodwin:1994jh} for the first few $S$- and $P$-wave channels.
The spin-triplet operators are normalized so that, if one makes the
replacement $\sigma_i \otimes \sigma_j \to \delta_{ij}(1 \otimes 1)$,
they become $(2j+1)/(2l+1)$ times the corresponding spin-singlet
operators. The color-octet operators are normalized so that, if one
makes the replacement $T^a \otimes T^a \to 1 \otimes 1$, they become the
corresponding color-singlet operators. The $P$-wave operators are
normalized so that, if one makes the replacement $(-\mbox{$\frac{i}{2}$}
\tensor{D}_i) \otimes (-\mbox{$\frac{i}{2}$} \tensor{D}_j) \to
\delta_{ij}(1 \otimes 1)$, they become  $(2j+1)/(2s+1)$  times the
corresponding $S$-wave operators. Given these normalization
conventions, we find that the CEM relation (\ref{nrqcd-cem-me}) implies
that the standard 
$S$-wave and $P$-wave NRQCD matrix elements are related by
\begin{equation}
\langle {\cal O}_n^H \rangle =
\frac{3(2j+1)}{(2l+1)(2l+3)} C_n k_{\rm max}^{2l}
\langle {\cal O}_1^H({}^1S_0) \rangle,
\label{me-cem}
\end{equation}
where $C_n=1$ or $(N_c^2-1)/(2N_c)=4/3$ if ${\cal O}_n^H$ is a color-singlet
or color-octet operator, respectively. 
To be more specific, the CEM
relation (\ref{nrqcd-cem-me}) implies that the $S$-wave matrix elements
are all related by simple group theory factors:
\begin{eqnarray}
\langle {\cal O}_1^H({}^3S_1) \rangle &=&
3 \langle {\cal O}_1^H({}^1S_0) \rangle,
\\
\langle {\cal O}_8^H({}^1S_0) \rangle &=&
\frac{4}{3} \langle {\cal O}_1^H({}^1S_0) \rangle,
\\
\langle {\cal O}_8^H({}^3S_1) \rangle &=&
4 \langle {\cal O}_1^H({}^1S_0) \rangle.
\end{eqnarray}
The $P$-wave matrix elements also include a factor $k_{\rm max}^2$ 
from the integral over the relative momentum:
\begin{eqnarray}
\langle {\cal O}_1^H({}^1P_1) \rangle &=&
\frac{3}{5} k_{\rm max}^2 \langle {\cal O}_1^H({}^1S_0) \rangle,
\\
\langle {\cal O}_1^H({}^3P_j) \rangle &=&
\frac{2j+1}{5} k_{\rm max}^2 \langle {\cal O}_1^H({}^1S_0) \rangle,
\\
\langle {\cal O}_8^H({}^1P_1) \rangle &=&
\frac{4}{5} k_{\rm max}^2 \langle {\cal O}_1^H({}^1S_0) \rangle,
\\
\langle {\cal O}_8^H({}^3P_j) \rangle &=&
\frac{4(2j+1)}{15} k_{\rm max}^2 \langle {\cal O}_1^H({}^1S_0) \rangle.
\end{eqnarray}

\section{Analysis of $\bm{S}$-wave Charmonium Production}
\label{sec:s-wave}

In the production of $S$-wave charmonium at the Tevatron
with transverse momentum $p_T > 5$ GeV, it is known phenomenologically
that the most important NRQCD matrix elements for $H=J/\psi$ or
$\psi(2S)$ are the color-octet matrix element $\langle{\cal
O}_8^H({}^3S_1)\rangle$ and a specific linear combination of color-octet
matrix elements $\langle{\cal O}_8^H({}^3P_0)\rangle$ and $\langle{\cal
O}_8^H({}^1S_0)\rangle$:
\begin{equation}
M_r^H=(r/m^2)\langle{\cal O}_8^H({}^3P_0)\rangle
+\langle{\cal O}_8^H({}^1S_0)\rangle,
\end{equation}
where $r\approx 3$.
Let us examine the ratio of these matrix elements
\begin{equation}
R^H=\frac{M_r^H}{\langle{\cal O}_8^H({}^3S_1)\rangle},
\label{r_h-defn}
\end{equation}
where $H$ stands for $J/\psi$ or $\psi(2S)$.
The relation (\ref{me-cem}) yields the CEM ratio
\begin{equation}
R^H_{\rm CEM}=\frac{M_r^H}{\langle{\cal O}_8^H({}^3S_1)\rangle}=
\frac{r}{15} \; \frac{k_{\rm max}^2}{m^2}+ \frac{1}{3}.
\label{r-S-wave}
\end{equation}
The velocity-scaling rules of NRQCD predict that all
of the matrix elements in both the numerator and denominator of
Eq.~(\ref{r_h-defn}) scale as $v^4$ relative to $\langle{\cal
O}_1^H({}^3S_1)\rangle$. Hence, the ratio in Eq.~(\ref{r_h-defn}) is
predicted to scale as $v^0$. Since $k_{\rm max}$ scales as $mv$, 
the second term in the CEM ratio in Eq.~(\ref{r-S-wave}) satisfies this
scaling relation, but the first term does not.

\subsection{Analysis of Tevatron Data on $\bm{J/\psi}$ Production}

We now use the relation (\ref{me-cem}) to test the validity of the CEM
in comparisons with data. In Table~\ref{tab:J/psi}, we show values of
$R^{J/\psi}$ for several different sets of NRQCD matrix elements that
have been extracted, under varying assumptions, from the
transverse-momentum distribution of $J/\psi$'s produced at the Tevatron.
The matrix elements were taken from the compilation of
Ref.~\cite{Kramer:2001hh}. We also show the CEM values
$R^{J/\psi}_{\rm CEM}$, taking $k^2_{\rm
max}/m^2=(m_D^2-m_c^2)/m_c^2\approx 0.54$.
\begin{table}
\caption{Values of $R^{J/\psi}$, as defined in Eq.~(\ref{r_h-defn}), in
the NRQCD factorization approach and in the CEM. The column labeled
``$R^{J/\psi}$'' gives phenomenological values of $R^{J/\psi}$ from
various extractions of the NRQCD matrix elements from the
CDF data \cite{Abe:1997yz}. The column labeled ``Reference'' gives the
reference for each extraction, and the column labeled ``PDF'' gives the
parton distribution that was used in the extraction. The headings ``LO
collinear factorization,'' ``parton-shower radiation,'' and ``$k_T$
smearing'' refer to the method that was used to compute the NRQCD
factorization prediction. The column labeled ``$R_{\rm CEM}^{J/\psi}$''
gives the CEM ratios from Eq.~(\ref{r-S-wave}) for
the values of $r$ and $m_c$ that were used in the NRQCD
extractions of $R^{J/\psi}$.
} 
\label{tab:J/psi}
\begin{ruledtabular}
\begin{tabular}{c|cc|ccccc}
Reference                & 
\multicolumn{2}{c|}{PDF} & 
$R^{J/\psi}$             &
$R^{J/\psi}_{\rm CEM}$        &
$r$ &
$m_c$~(GeV)&
$\langle k_T \rangle$ \mbox{(GeV)} \\
\hline
\hline
\multicolumn{8}{c}{\mbox{LO collinear factorization}} \\ 
\hline \cite{Cho:1996ce} & \multicolumn{2}{c|}{MRS(D0)~\cite{Martin:1993zi}} & 
10 $\pm$ 4 &  0.44 & 3 & 1.48&\\
\hline                   & 
\multicolumn{2}{c|}{\mbox{CTEQ4L~\cite{Lai:1997mg}}} & 
4.1 $\pm$ 1.2~${}^{+3.6}_{-1.3}$ &  & &&\\
 \cite{Beneke:1997yw}    & 
\multicolumn{2}{c|}{\mbox{GRV-LO(94)~\cite{Gluck:1995uf}}} & 
3.5 $\pm$ 1.1~${}^{+1.6}_{-0.9}$ & 0.46 & 3.5 & 1.5&\\
                         & 
\multicolumn{2}{c|}{\mbox{MRS(R2)~\cite{Martin:1996as}}} &
7.8 $\pm$ 1.9~${}^{+8.0}_{-2.8}$ &  &  && \\
\hline                   & 
\multicolumn{2}{c|}{\mbox{MRST-LO(98)~\cite{Martin:1998sq}}} & 
20 $\pm$ 4 &  &  && \\
\raisebox{2ex}[-2ex]{\cite{Braaten:2000qk}} & 
\multicolumn{2}{c|}{\mbox{CTEQ5L~\cite{Lai:1999wy}}} & 
17 $\pm$ 4  & \raisebox{2ex}[-2ex]{0.46}
            & \raisebox{2ex}[-2ex]{3.4}
            & \raisebox{2ex}[-2ex]{1.5}& \\
\hline\hline
\multicolumn{8}{c}{\mbox{parton-shower radiation}} \\
 \hline &
 \multicolumn{2}{c|}{\mbox{CTEQ2L}~\cite{Tung:1994ua}} & 
1.4 $\pm$ 0.3 &  &  & \\
 \cite{Sanchis-Lozano:2000um} &
\multicolumn{2}{c|}{\mbox{MRS(D0)}~\cite{Martin:1993zi}} & 
1.9 $\pm$ 0.6 & 0.44 & 3 & 1.48&\\
                              & 
\multicolumn{2}{c|}{\mbox{GRV-HO(94)}~\cite{Gluck:1995uf}} & 
0.49 $\pm$ 0.11 &  & & \\
\hline 
\cite{Kniehl:1999qy} &
\multicolumn{2}{c|}{\mbox{CTEQ4M}~\cite{Lai:1997mg}} & 
2.1 $\pm$ 0.8 &  0.45 & 3.5 & 1.55&\\[0.5mm]
\hline \hline
 \multicolumn{8}{c}{\mbox{$k_T$ smearing}} \\ 
\hline 
&&&5.7 $\pm$ 1.6 &  &  & & 1.0\\
 \raisebox{2ex}[-2ex]{\cite{Petrelli:2000rh}} &
 \raisebox{2ex}[-2ex]{CTEQ4M~\cite{Lai:1997mg}} & & 
2.6 $\pm$ 0.9 & \raisebox{2ex}[-2ex]{0.46}   
              & \raisebox{2ex}[-2ex]{3.5} 
              & \raisebox{2ex}[-2ex]{1.5}&1.5\\ 
\hline 
&&&6.3 $\pm$ 1.7 &  &  &  & 0.7\\ 
\raisebox{2ex}[-2ex]{\cite{Sridhar:1998rt}}
 & \raisebox{2ex}[-2ex]{\mbox{MRS(D$'_-$)~\cite{Martin:1993zi}}} & & 
4.7 $\pm$ 1.2 & \raisebox{2ex}[-2ex]{$\approx 0.44$}
              & \raisebox{2ex}[-2ex]{3} 
              & \raisebox{2ex}[-2ex]{$\approx$ 1.5} &1.0\\[0.5mm]
\end{tabular}
\end{ruledtabular}
\end{table}
Several sets of matrix elements were extracted by making use of NRQCD
short-distance coefficients that were calculated at leading order in
$\alpha_s$ and under the assumption of standard collinear factorization.
For these sets of matrix elements, $R^{J/\psi}$ is much larger than 
$R^{J/\psi}_{\rm CEM}$. Multiple gluon radiation, as
modeled by parton-shower Monte Carlos, tends to increase the partonic
cross section more at smaller values of $p_T$ than at larger values of
$p_T$. Since the contribution of $\langle{\cal
O}_8^{J/\psi}({}^3S_1)\rangle$ dominates that of $M_r^{J/\psi}$ at large
$p_T$, while the contribution of $M_r^{J/\psi}$ is the more important
one at small $p_T$, the effect of parton showering is to decrease the
size of $R^{J/\psi}$. Hence, the addition of parton showering to the
leading-order calculation of the NRQCD short-distance coefficients
brings the ratio $R^{J/\psi}$ for the extracted values of the matrix
elements into better agreement with the $R^{J/\psi}_{\rm CEM}$. However, 
the extraction that is based on the more recent CTEQ(4M) parton
distributions is still in significant disagreement with 
$R^{J/\psi}_{\rm CEM}$.
Surprisingly, $k_T$ smearing does not decrease the size of 
$R^{J/\psi}$ as much as parton showering, and there is a substantial 
disagreement between $R^{J/\psi}_{\rm CEM}$ and the values of $R^{J/\psi}$ that 
are obtained by using $k_T$ smearing. 

Now let us compare the predictions of the CEM and NRQCD factorization 
directly with the CDF data \cite{Abe:1997yz}. 

The CEM predictions are from a calculation by Vogt \cite{vogt} that
makes use of the order-$\alpha_s^3$ cross section for production of a
$Q\bar Q$ pair \cite{Mangano:1992kq}. Details of this calculation are
given in Ref.~\cite{Brambilla:2004wf}. The CEM factors $F_H$ in
Eq.~(\ref{prod-cem}) were fixed by comparison with fixed-target data.
The charm-quark mass $m_c$ was also tuned to optimize the fits. The
factorization and renormalization scales were chosen to be $\mu \propto
m_T=\sqrt{m_c^2+p_T^2}$, where $p_T$ is the sum of the transverse
momenta of the $Q$ and the $\bar Q$. In our comparisons, we make use of
the parameter sets labeled ``$\psi 1$'' and ``$\psi 4$'' in
Ref.~\cite{Brambilla:2004wf}, which correspond, respectively, to
$\mu=2m_T$, $m_c=1.2$~GeV, MRST98~HO \cite{Martin:1998sq} parton
distributions and $\mu=m_T$, $m_c=1.3$~GeV, GRV98~HO \cite{Gluck:1998xa}
parton distributions. It is for these two parameter sets that CEM
predictions of the $p_T$ distributions of charmonia produced at the
Tevatron are available.

The NRQCD predictions were generated from modified versions of computer
codes created by Maltoni, Mangano, and Petrelli \cite{MMP}. The codes
compute the order-$\alpha_s^3$ quarkonium production cross sections
\cite{Petrelli:1997ge} and the standard
Dokshitzer-Gribov-Lipatov-Altarelli-Parisi (DGLAP) evolution of the
fragmentation contribution to the evolution of a $Q\bar Q$ pair in a
${}^3S_1$ color-octet state into a quarkonium. This fragmentation
contribution is the dominant contribution at large $p_T$. We chose
$m_c=1.5$~GeV and took the factorization and renormalization scales to be
$\mu=m_T=\sqrt{m_c^2+p_T^2}$. Our calculations made use of the same parton
distributions as the CEM predictions, namely the MRST98~HO
\cite{Martin:1998sq} parton distributions and the GRV98~HO
\cite{Gluck:1998xa} parton distributions.

\begin{figure}
\begin{tabular}{cc}
\includegraphics[width=8cm]{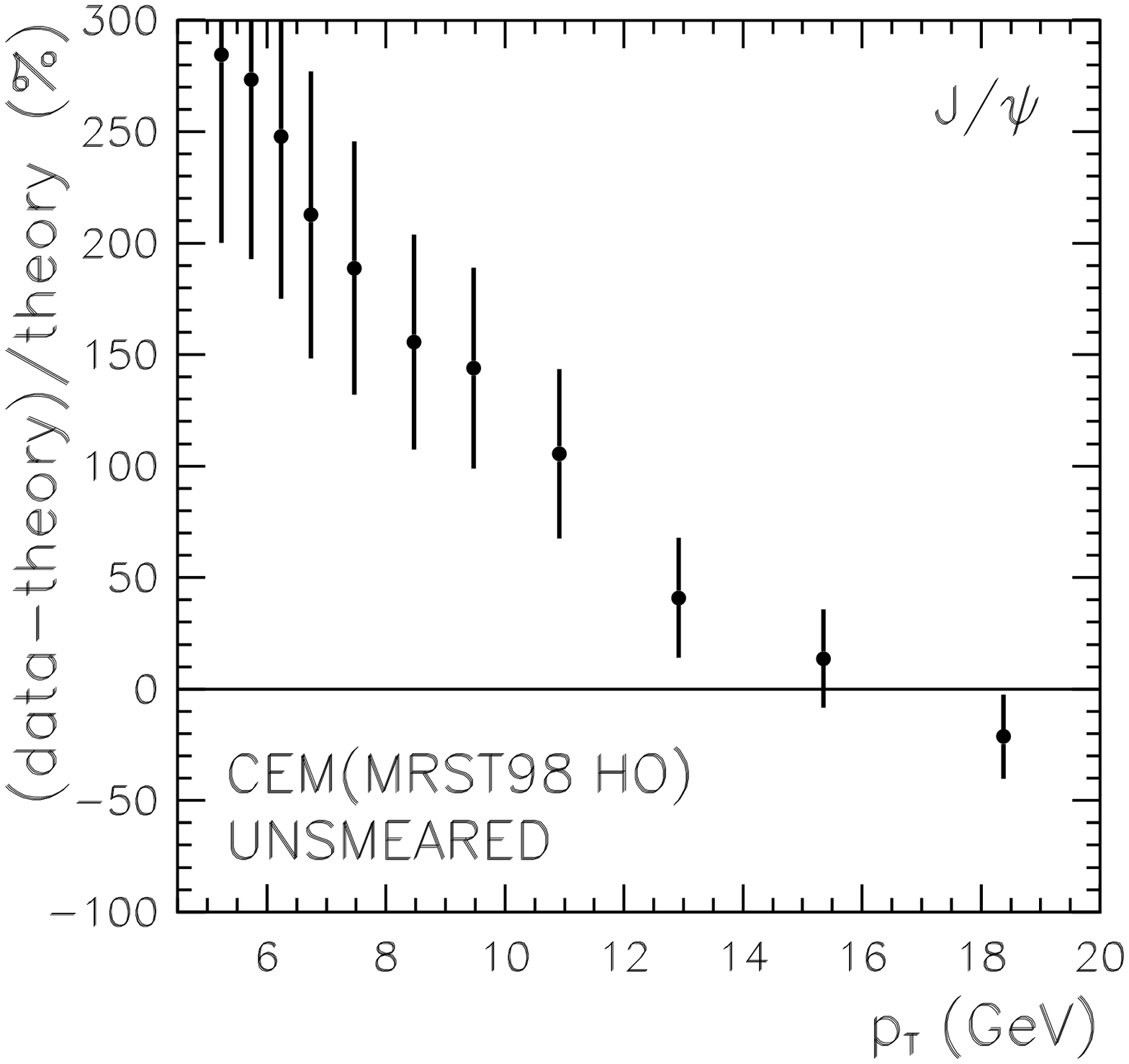}& 
\includegraphics[width=8cm]{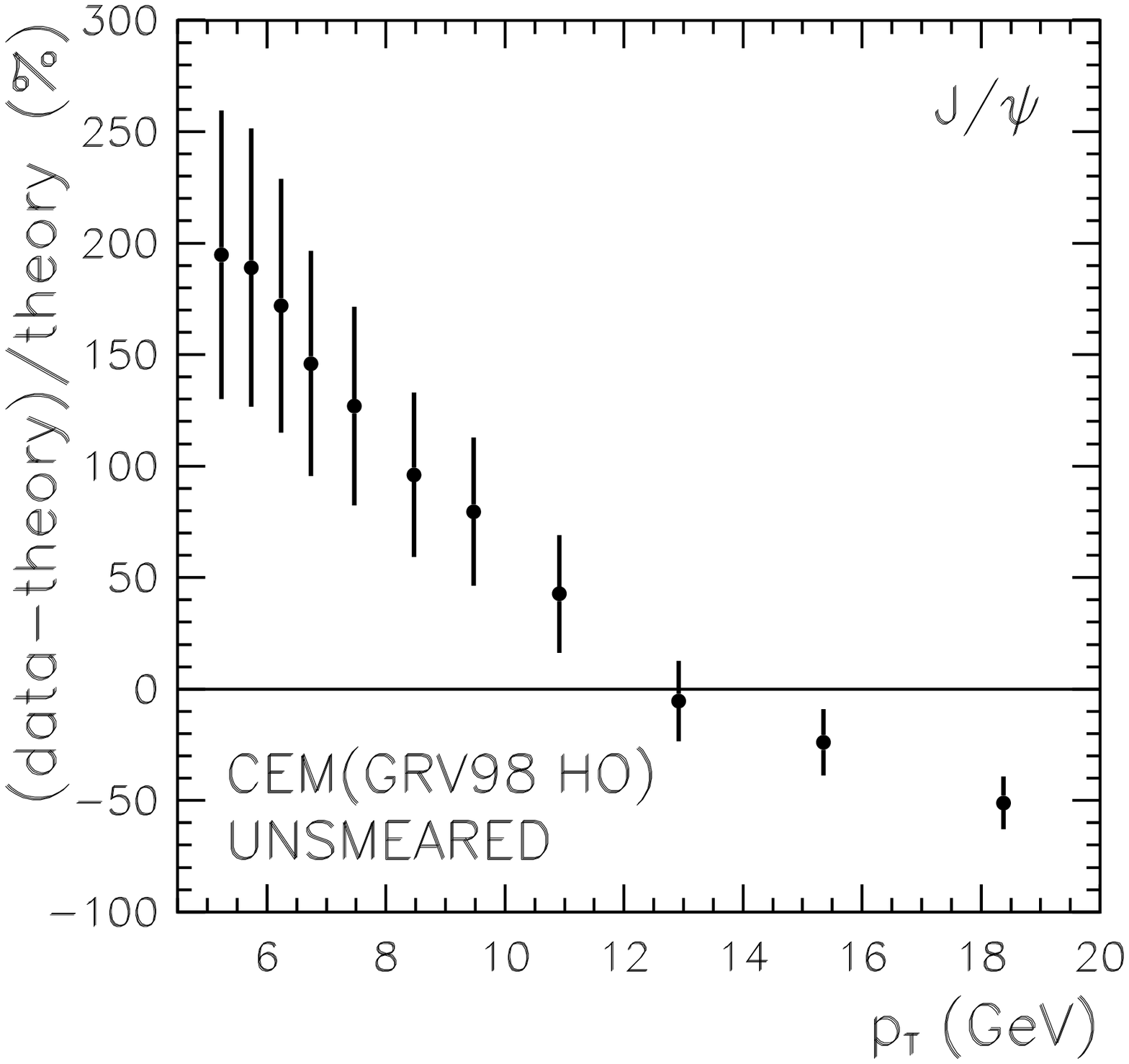}\\
\includegraphics[width=8cm]{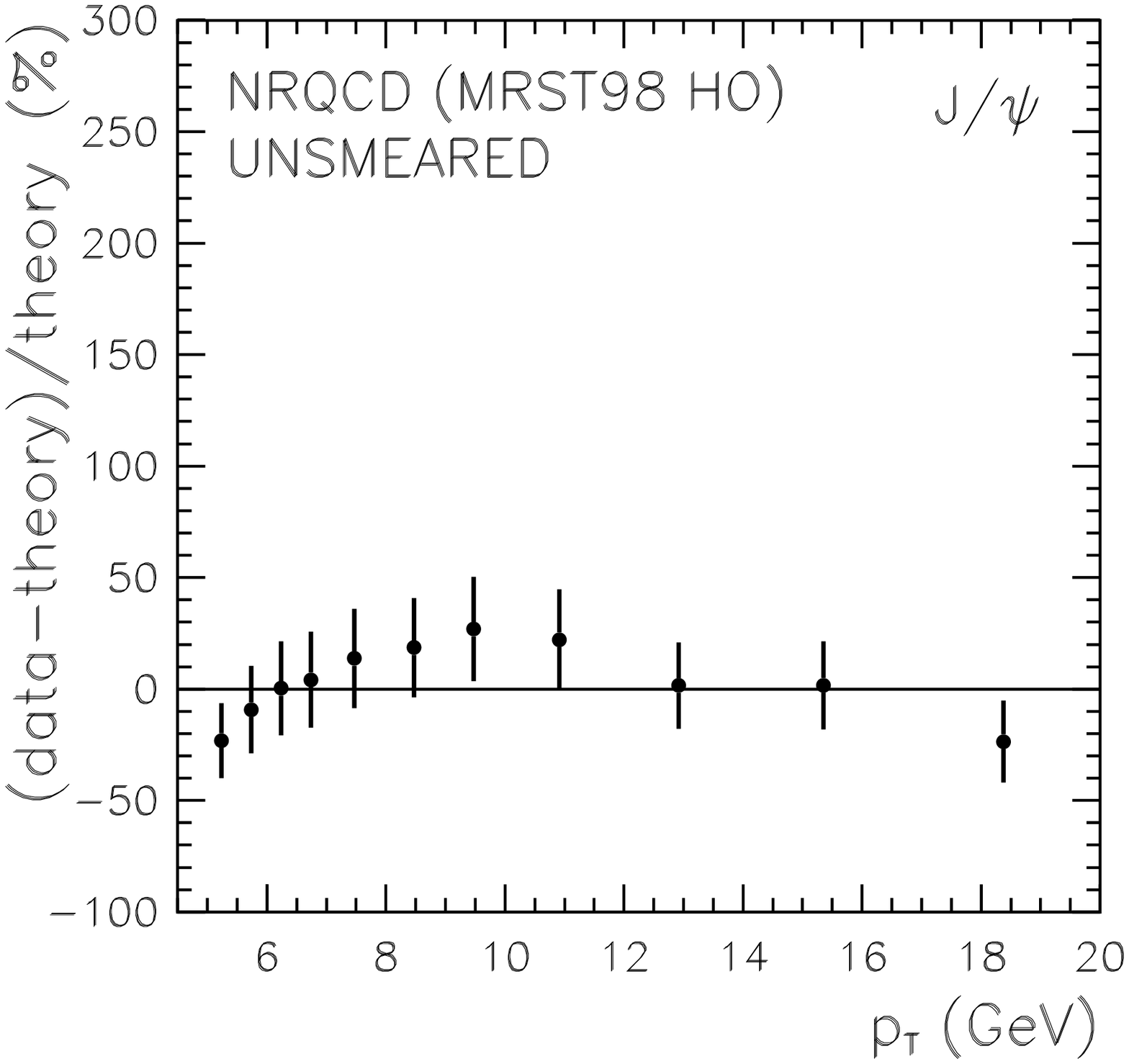}& 
\includegraphics[width=8cm]{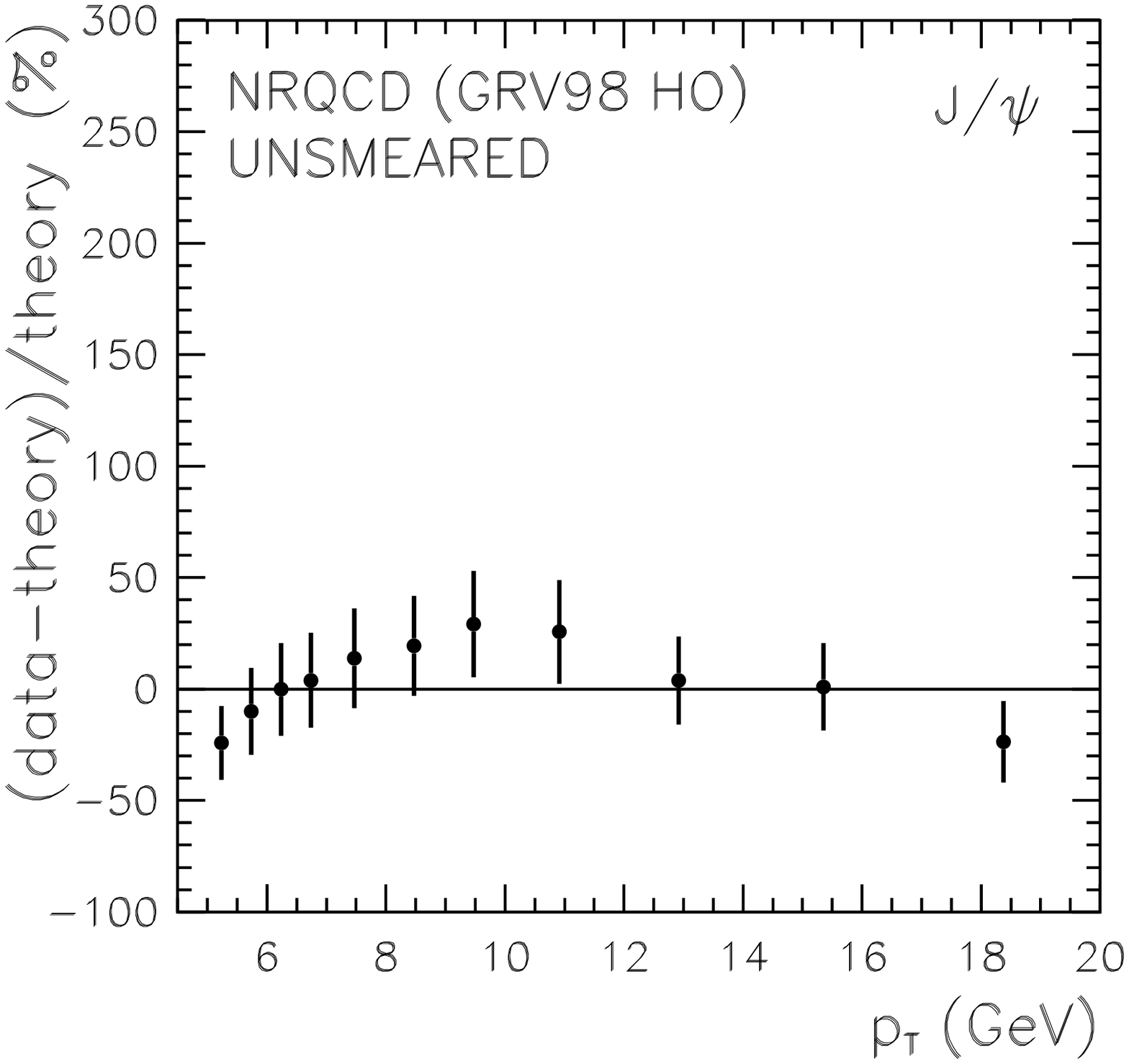}  
\end{tabular}
\caption{$J/\psi$ production: $({\rm data}-{\rm theory})/{\rm theory}$.
The data are from the measurements of the CDF collaboration
\cite{Abe:1997yz}. The upper figures are for the CEM predictions and
the lower figures are for the NRQCD factorization predictions. The
theoretical predictions for the left-hand and right-hand figures are
based on the MRST98~HO \cite{Martin:1998sq} and the GRV98~HO
\cite{Gluck:1998xa} parton distributions, respectively.}
\label{fig:psi-unsmeared}
\end{figure}

Plots of $({\rm data}-{\rm theory})/{\rm theory}$ are shown for the CEM
and NRQCD factorization predictions in Fig.~\ref{fig:psi-unsmeared}. 
There is a substantial disagreement between the CEM predictions and the
data. The normalizations of the CEM predictions are too small, and the
slopes are relatively too positive. This discrepancy in the slopes is
consistent with the fact that the CEM relation (\ref{r-S-wave})
over-estimates the size of $\langle{\cal O}_8^{J/\psi}({}^3S_1)\rangle$
relative to $M_r^{J/\psi}$. The NRQCD factorization predictions are in much
better agreement with the data than the CEM predictions. Even if one were
to adjust the normalizations of the CEM predictions to improve the fits, 
they would still be unsatisfactory, owing to the differences in slope 
between the CEM predictions and the data.

A simple phenomenological model for the effects of multiple gluon
emission on the theoretical predictions is $k_T$ smearing. In $k_T$
smearing, the colliding partons are given Gaussian distributions in the
intrinsic transverse momentum, with a width that is treated as a
phenomenological parameter. A particular version of this model that has
been used in comparing the CEM predictions with the experimental data
\cite{vogt,Brambilla:2004wf} attempts to account for multiple gluon
emission from the two initial-state partons by adding two 
transverse-momentum ``kicks'' to the quarkonium momentum. The direction of
each momentum kick is symmetrically distributed over the $4\pi$ solid
angle, and the magnitude $k_T$ of each momentum kick is distributed
as
\begin{equation}
g(k_T)=\frac{1}{\pi\langle k_T^2\rangle}\exp(-k_T^2/\langle 
k_T^2\rangle),
\end{equation}
where $\langle k_T^2\rangle$ is a phenomenological parameter. In the
case of the CEM predictions with $k_T$ smearing, $\langle k_T^2\rangle$
has been tuned to the value $\langle k_T^2\rangle=2.5~\hbox{GeV}^2$ in
order to obtain the best fit to the CDF $J/\psi$ data
\cite{vogt,Brambilla:2004wf}. This same value was also used in making
CEM predictions for $\psi(2S)$ and $\chi_c$ production. For purposes of
comparison, we also take $\langle k_T^2\rangle=2.5~\hbox{GeV}^2$ when we
apply $k_T$ smearing to the NRQCD factorization predictions for
charmonium production.

\begin{figure}
\begin{tabular}{cc}
\includegraphics[width=8cm]{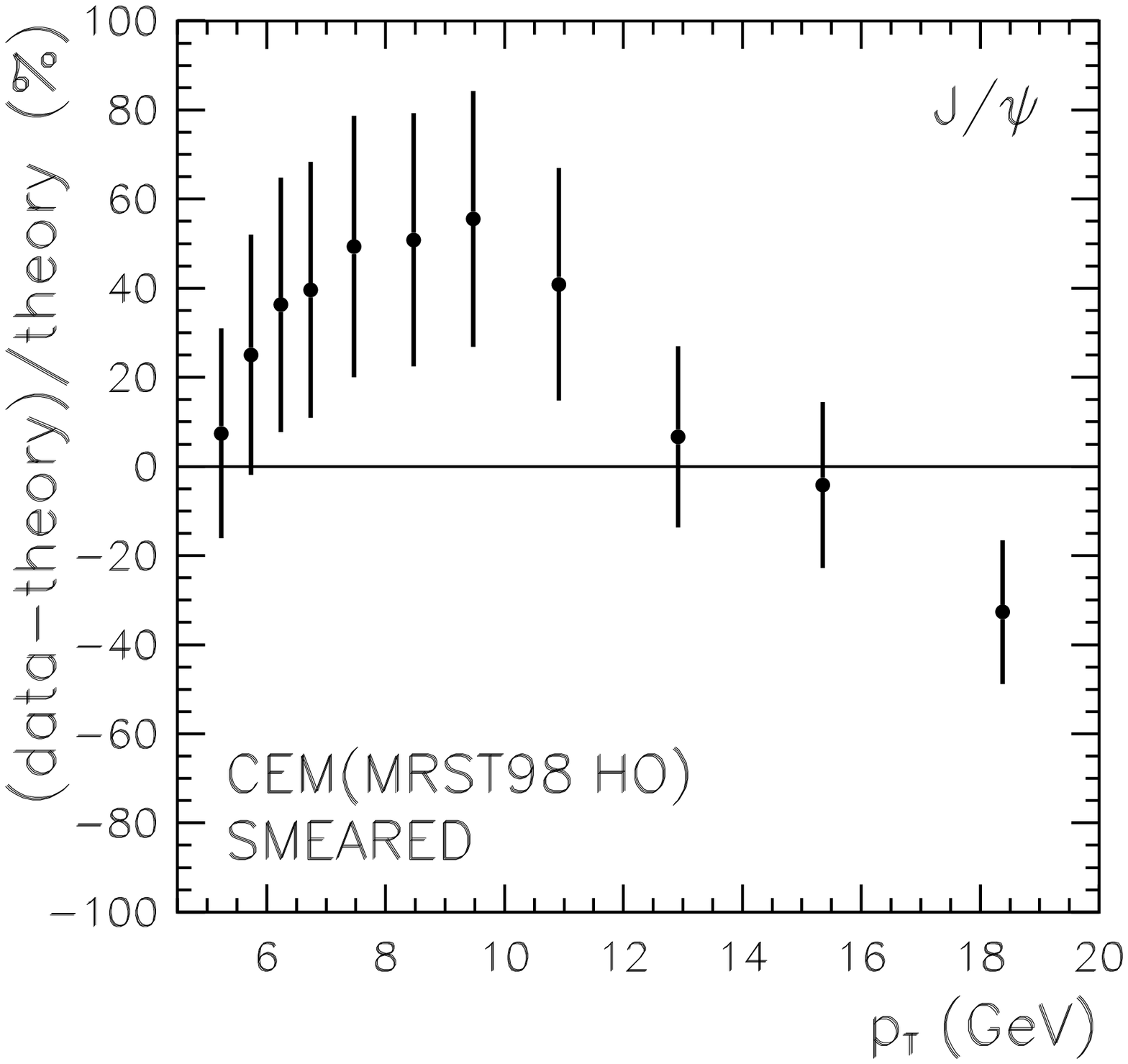}& 
\includegraphics[width=8cm]{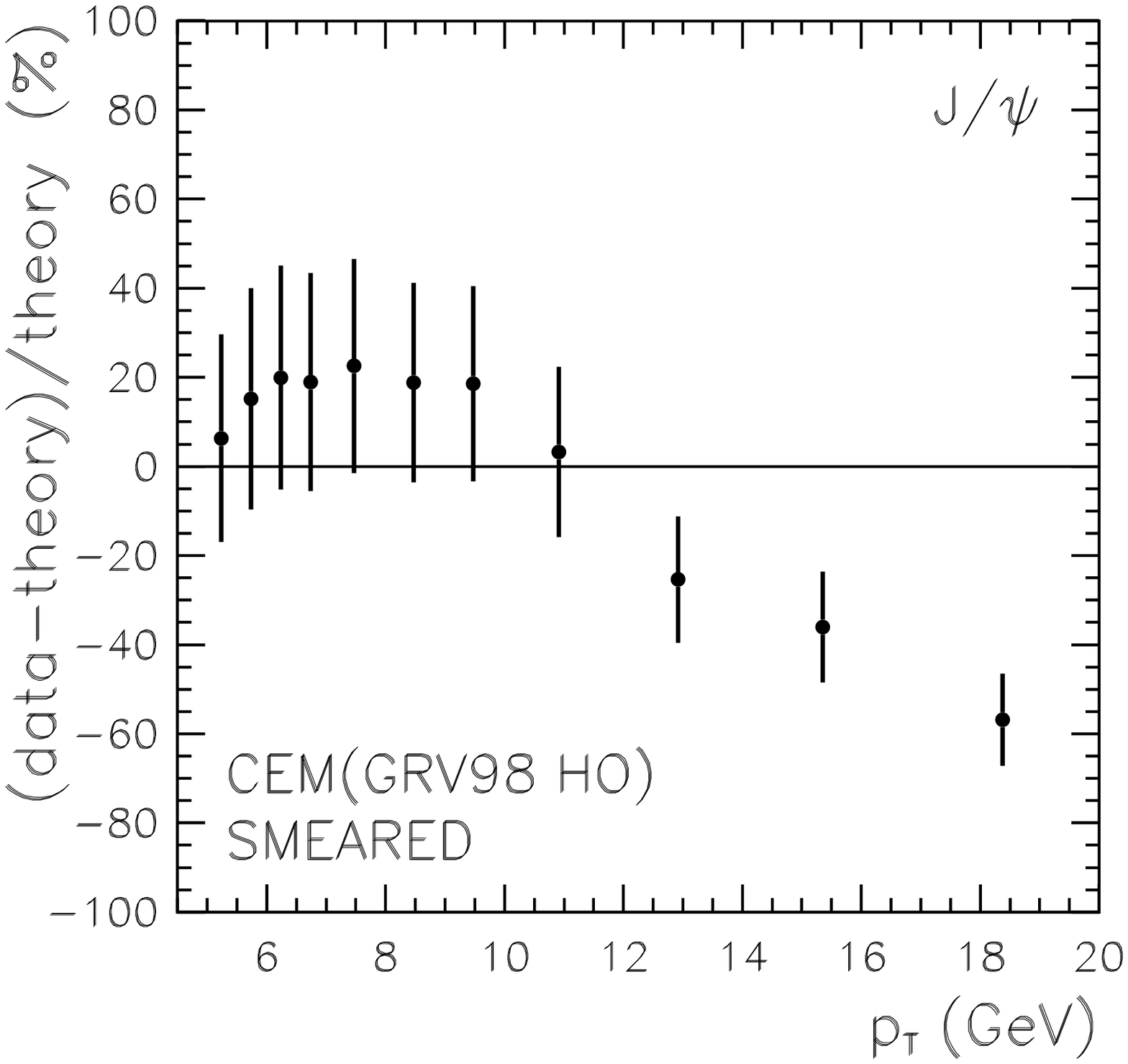}\\
\includegraphics[width=8cm]{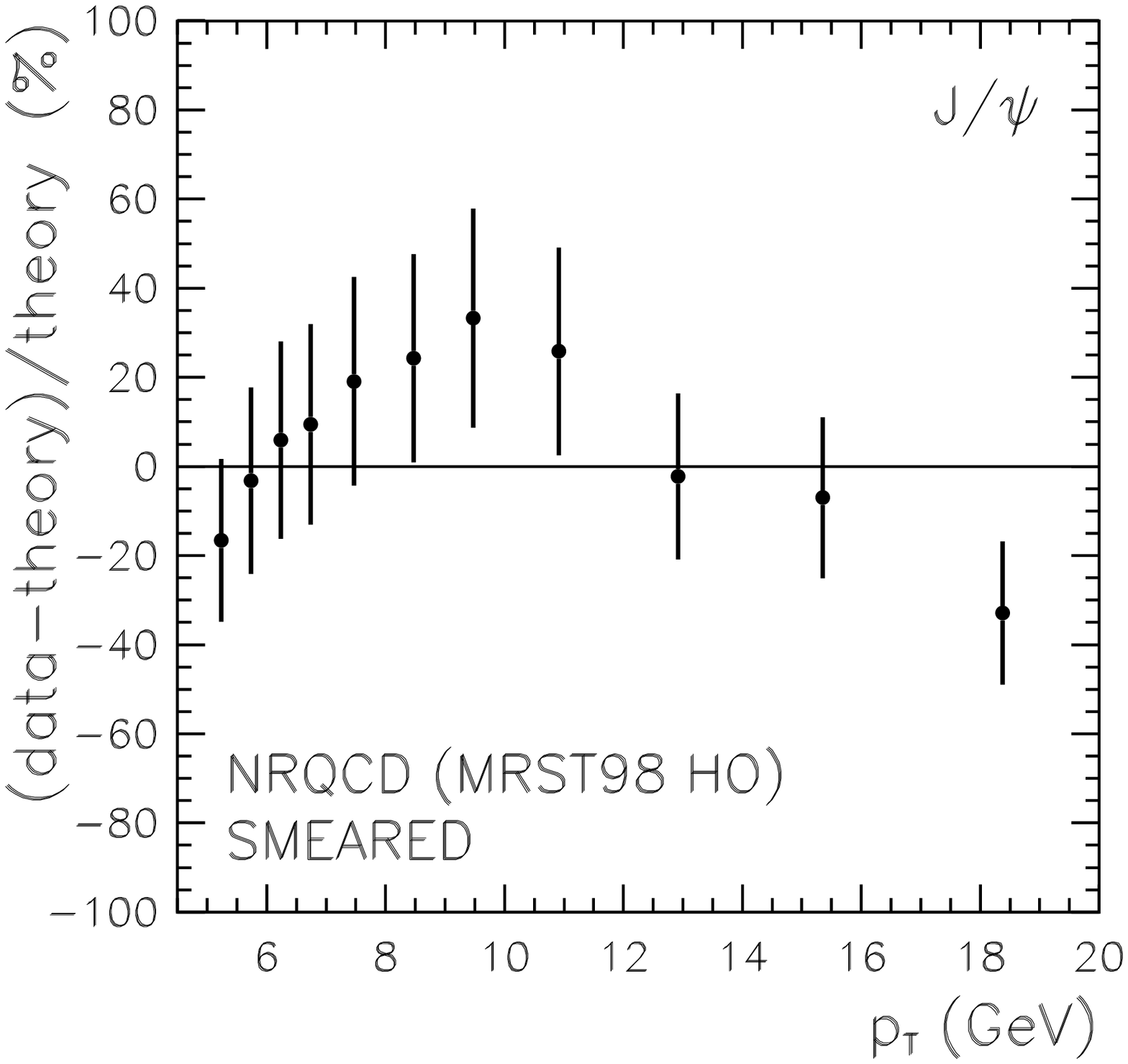}& 
\includegraphics[width=8cm]{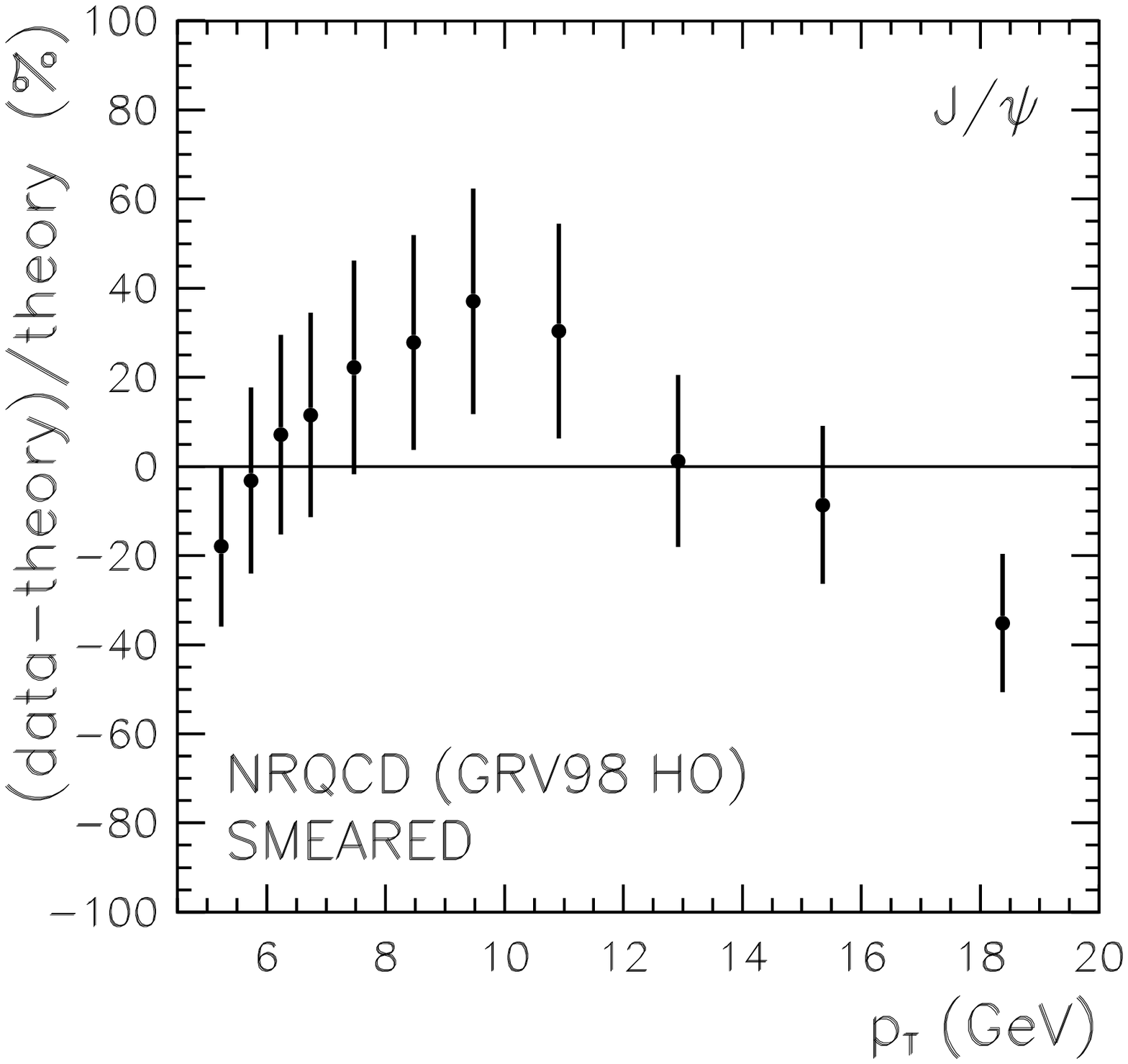}  
\end{tabular}
\caption{$J/\psi$ production: $({\rm data}-{\rm theory})/{\rm theory}$,
with $k_T$ smearing of the theory, as described in the text. The data
are from the measurements of the CDF collaboration \cite{Abe:1997yz}.
The upper figures are for the CEM predictions and the lower figures are
for the NRQCD factorization predictions. The theoretical predictions
for the left-hand and right-hand figures are based on the MRST98~HO
\cite{Martin:1998sq} and the GRV98~HO \cite{Gluck:1998xa} parton
distributions, respectively.}
\label{fig:psi-smeared}
\end{figure}

Plots of $({\rm data}-{\rm theory})/{\rm theory}$ for the CEM and NRQCD
factorization predictions with $k_T$ smearing are shown in
Fig.~\ref{fig:psi-smeared}.
The $k_T$-smearing procedure substantially improves both the slope
and normalization of the CEM fits to the data and slightly worsens
the NRQCD factorization fits to the data.\footnote{The effects of
$k_T$ smearing on predictions for $J/\psi$ production cross sections at
the Tevatron have also been studied by Sridhar, Martin, and Stirling
\cite{Sridhar:1998rt} and Petrelli \cite{Petrelli:2000rh}. These studies
made use of somewhat smaller values of $\langle k_T^2\rangle$
than in the present work. They also concluded that the quality of the NRQCD
factorization fits to the CDF data is little affected by $k_T$
smearing.} The improvement of the slopes of the CEM predictions
with $k_T$ smearing is to be expected since, generally, the effect
of $k_T$ smearing is to increase the cross section considerably at
moderate values of $p_T$ and increase the cross section by a smaller
amount at high $p_T$. Nevertheless, the $k_T$-smeared CEM predictions
yield poor fits to the data, as the slopes are still too positive
relative to the data. This is consistent with the fact that the CEM
relation (\ref{r-S-wave}) over-estimates the size of $\langle{\cal
O}_8^{J/\psi}({}^3S_1)\rangle$ relative to $M_r^{J/\psi}$, even for
$k_T$-smeared extractions of the matrix elements.

A compilation of values of matrix elements,  values of $R^{J/\psi}$, 
and chi-squared per degree of freedom ($\chi^2/{\rm d.o.f.}$) from the 
NRQCD factorization fits and CEM fits to the $J/\psi$ data is given in
Table~\ref{tab:psi-fits}. In the NRQCD factorization fits, the degrees
of freedom are reduced by two, owing to the two NRQCD matrix elements
that are varied in the fits. The unsmeared CEM fits have no free
parameters, as the overall normalization is fixed by comparison with the
fixed-target data. The smeared CEM fits to the $J/\psi$ data have one
free parameter, $\langle k_T^2\rangle$, which is then held constant in
fits to the $\psi(2S)$ and $\chi_c$ data. The values of $R^{J/\psi}$
in Table~~\ref{tab:psi-fits} are much greater than $R_{\rm
CEM}^{J/\psi}\approx 0.46$ in the fits without $k_T$ smearing and somewhat
greater than $R_{\rm CEM}^{J/\psi}$ in the fits with $k_T$ smearing.
The relative values of $R^{J/\psi}$ and $R_{\rm CEM}^{J/\psi}$ are
consistent with the large discrepancy between the slopes of the NRQCD
factorization and CEM fits without $k_T$ smearing and the smaller
discrepancy between the slopes of the NRQCD factorization and CEM fits
with $k_T$ smearing.
\begin{table}
\caption{Values of matrix elements, $R^{J/\psi}$, and $\chi^2/{\rm
d.o.f.}$ from the NRQCD factorization and CEM fits to the $J/\psi$ data.
In the NRQCD factorization fits, we set $\langle{\cal
O}^{J/\psi}_1({}^3S_1)\rangle$=1.16~GeV$^3$ and give the fitted values
of $\langle{\cal O}^{J/\psi}_8({}^3S_1)\rangle$ and
$M_{3.5}^{J/\psi}$.}
\label{tab:psi-fits}
\begin{ruledtabular}
\begin{tabular}{l|cccr}
PDF & $\langle{\cal O}^{J/\psi}_8({}^3S_1)\rangle$
&$M_{3.5}^{J/\psi}$&$R^{J/\psi}$
&$\chi^2/\textrm{d.o.f.}$   \\
&(GeV$^3\times 10^{-2}$)&(GeV$^3\times 10^{-2})$&&\\
\hline\hline
\multicolumn{5}{c}{NRQCD Factorization}\\
\hline
MRST98~HO          & 1.00 $\pm$ 0.22
                   & 8.83 $\pm$ 1.24
                   & 8.83 $\pm$ 2.27
                   & 7.16/(11$-$2)=0.80\\
GRV98~HO           & 1.02  $\pm$ 0.23
                   & 10.6  $\pm$ 1.42
                   & 10.4 $\pm$ 2.76
                   & 7.98/(11$-$2)=0.89\\
MRST98~HO (smeared) & 1.41 $\pm$ 0.13
                   & 0.41 $\pm$ 0.15
                   & 0.29 $\pm$ 0.11
                   & 10.28/(11$-$2)=1.14\\
GRV98~HO (smeared)  & 1.54 $\pm$ 0.14
                   & 0.49 $\pm$ 0.16
                   & 0.32 $\pm$ 0.11
                   &12.69/(11$-$2)=1.41\\
\hline\hline
\multicolumn{5}{c}{Color-Evaporation Model}\\
\hline
MRST98~HO           &&&& 89.18/11=8.11\\
GRV98~HO            &&&& 80.86/11=7.35\\
MRST98~HO (smeared) &&&& 20.78/(11$-$1)=2.08\\
GRV98~HO (smeared)  &&&& 45.70/(11$-$1)=4.57
\end{tabular}
\end{ruledtabular}
\end{table}

\subsection{Analysis of Tevatron Data on $\bm{\psi(2S)}$ Production}

Next let us examine the case of $\psi(2S)$ production. In
Table~\ref{tab:psi(2S)} we show values of $R^{\psi(2S)}$ that were
obtained from several different sets of NRQCD matrix elements that have
been extracted from the transverse-momentum distribution of $\psi(2S)$'s
produced at the Tevatron. Again, the matrix elements were taken from the
compilation of Ref.~\cite{Kramer:2001hh}. We also show the CEM
values $R_{\rm CEM}^{\psi(2S)}$ in Table~\ref{tab:psi(2S)}. As
in the analysis of the $J/\psi$ data, the values of $R^{\psi(2S)}$ lie
substantially above $R_{\rm CEM}^{\psi(2S)}$, except in the case of the
matrix elements that were extracted by making use of the GRV-HO(94)
parton distributions and parton-shower radiation in computing the NRQCD
factorization predictions.
\begin{table}
\caption{Values of $R^{\psi(2S)}$, as defined in Eq.~(\ref{r_h-defn}), in
the NRQCD factorization approach and in the CEM. As in
Table~\ref{tab:J/psi}, but for $\psi(2S)$.
}
\label{tab:psi(2S)}
\begin{ruledtabular}
\begin{tabular}{c|cc|cccc}
Reference                & 
\multicolumn{2}{c|}{PDF} & 
$R^{\psi(2S)}$      &
$R^{\psi(2S)}_{\rm CEM}$        &
$r$ &
$m_c$~(GeV)\\
\hline\hline
\multicolumn{7}{c}{\mbox{LO collinear factorization}} \\ 
\hline \cite{Cho:1996ce} & \multicolumn{2}{c|}{MRS(D0)~\cite{Martin:1993zi}} & 
 3.8 $\pm$ 1.5 & 0.44 & 3 & 1.48\\
\hline                   & 
\multicolumn{2}{c|}{\mbox{CTEQ4L~\cite{Lai:1997mg}}} & 
 4.1 $\pm$ 1.5~${}^{+3.4}_{-1.3}$ &  &  &\\
 \cite{Beneke:1997yw}    & 
\multicolumn{2}{c|}{\mbox{GRV-LO(94)~\cite{Gluck:1995uf}}} & 
 3.5 $\pm$ 1.3~${}^{+1.6}_{-0.9}$ &  0.46 & 3.5 & 1.5\\
                         & 
\multicolumn{2}{c|}{\mbox{MRS(R2)~\cite{Martin:1996as}}} &
 7.8 $\pm$ 2.3~${}^{+8.3}_{-2.8}$ &  &  &\\
\hline                   & 
\multicolumn{2}{c|}{\mbox{MRST-LO(98)~\cite{Martin:1998sq}}} & 
 3.1 $\pm$ 1.4 &  &  &\\
\raisebox{2ex}[-2ex]{\cite{Braaten:2000qk}} & 
\multicolumn{2}{c|}{\mbox{CTEQ5L~\cite{Lai:1999wy}}} & 
 2.1 $\pm$ 1.1 & \raisebox{2ex}[-2ex]{0.46}
               & \raisebox{2ex}[-2ex]{3.5}
               & \raisebox{2ex}[-2ex]{1.5} \\
\hline\hline
\multicolumn{7}{c}{\mbox{parton-shower radiation}} \\
 \hline &
 \multicolumn{2}{c|}{\mbox{CTEQ2L}~\cite{Tung:1994ua}} & 
 2.4 $\pm$ 0.8 &  &   &\\
 \cite{Sanchis-Lozano:2000um,Cano-Coloma:1997rn} &
\multicolumn{2}{c|}{\mbox{MRS(D0)}~\cite{Martin:1993zi}} & 
 2.5 $\pm$ 0.9 &  0.44 & 3 & 1.48\\
                              & 
\multicolumn{2}{c|}{\mbox{GRV-HO(94)}~\cite{Gluck:1995uf}} & 
 0.28 $\pm$ 0.35 &  &  & 
\end{tabular}
\end{ruledtabular}
\end{table}

\begin{figure}
\begin{tabular}{cc}
\includegraphics[width=8cm]{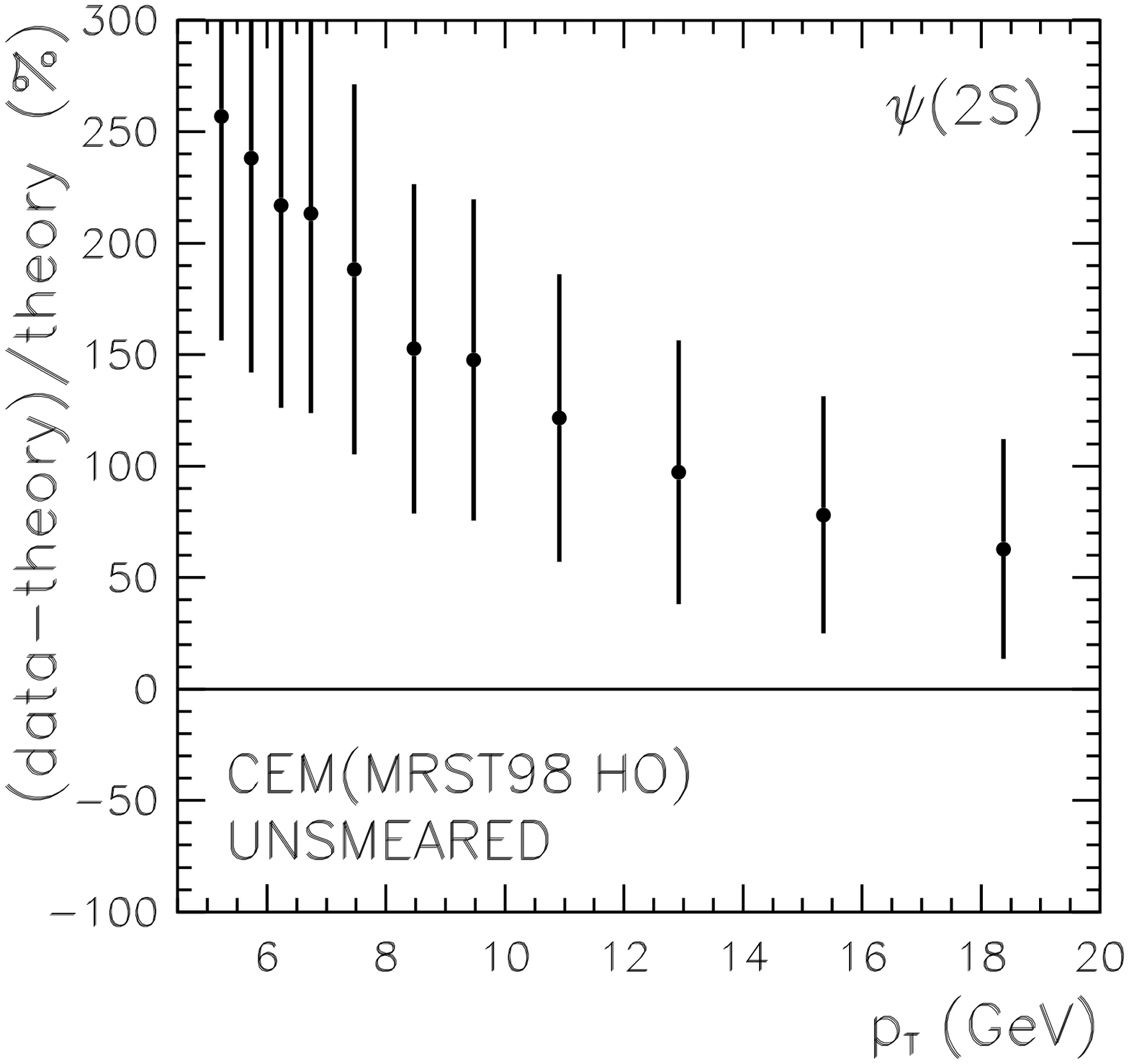}& 
\includegraphics[width=8cm]{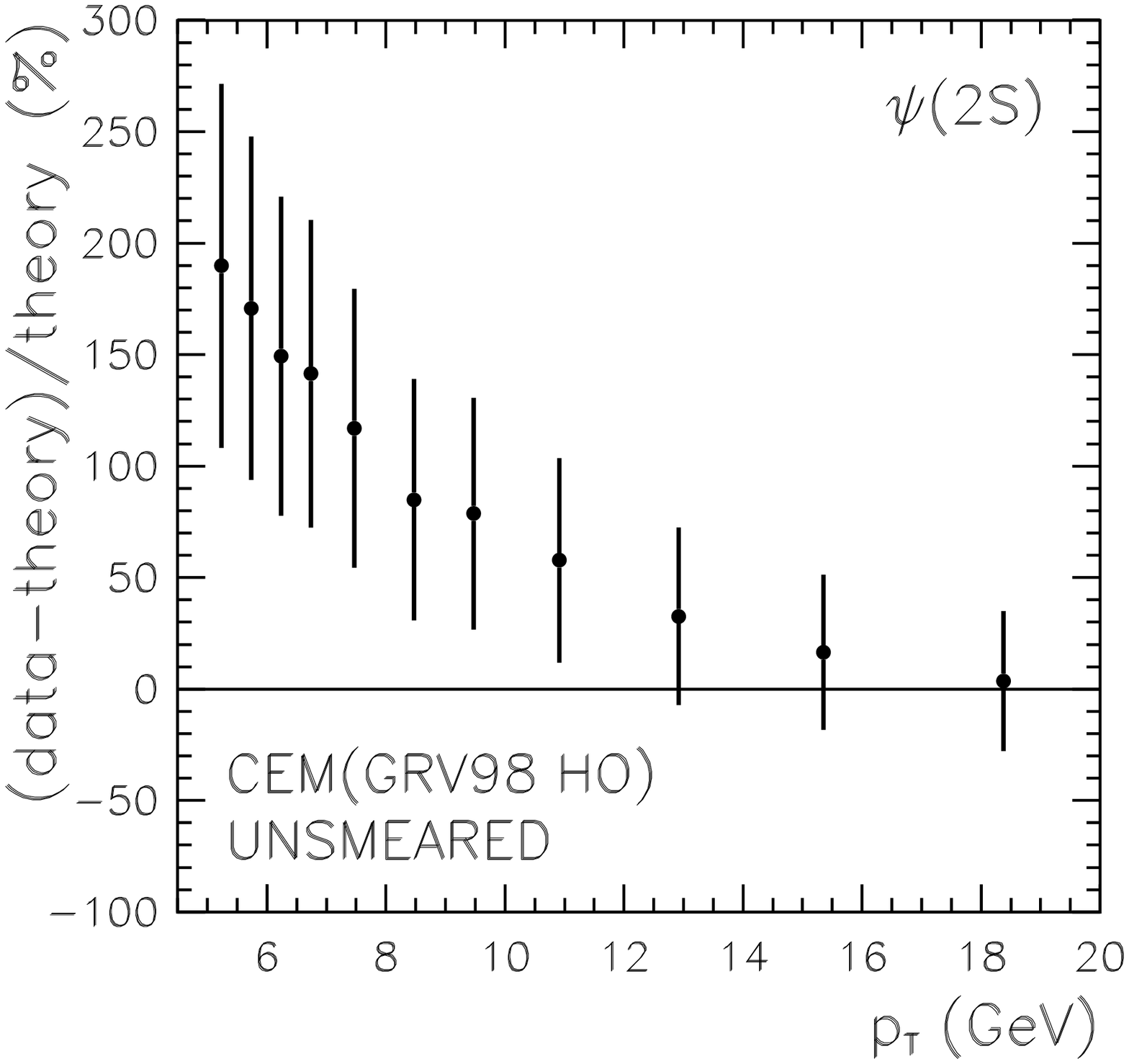}\\
\includegraphics[width=8cm]{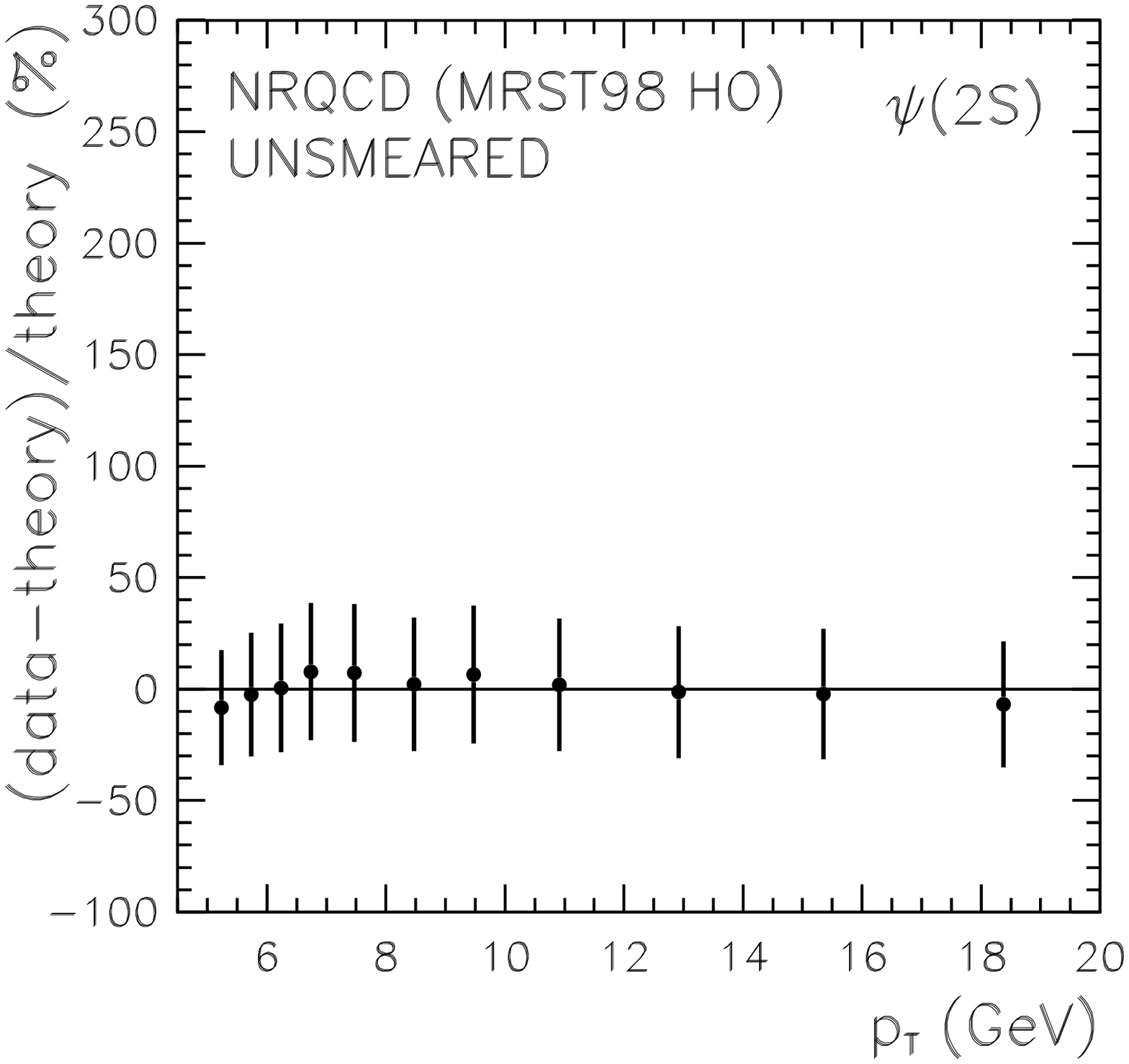}& 
\includegraphics[width=8cm]{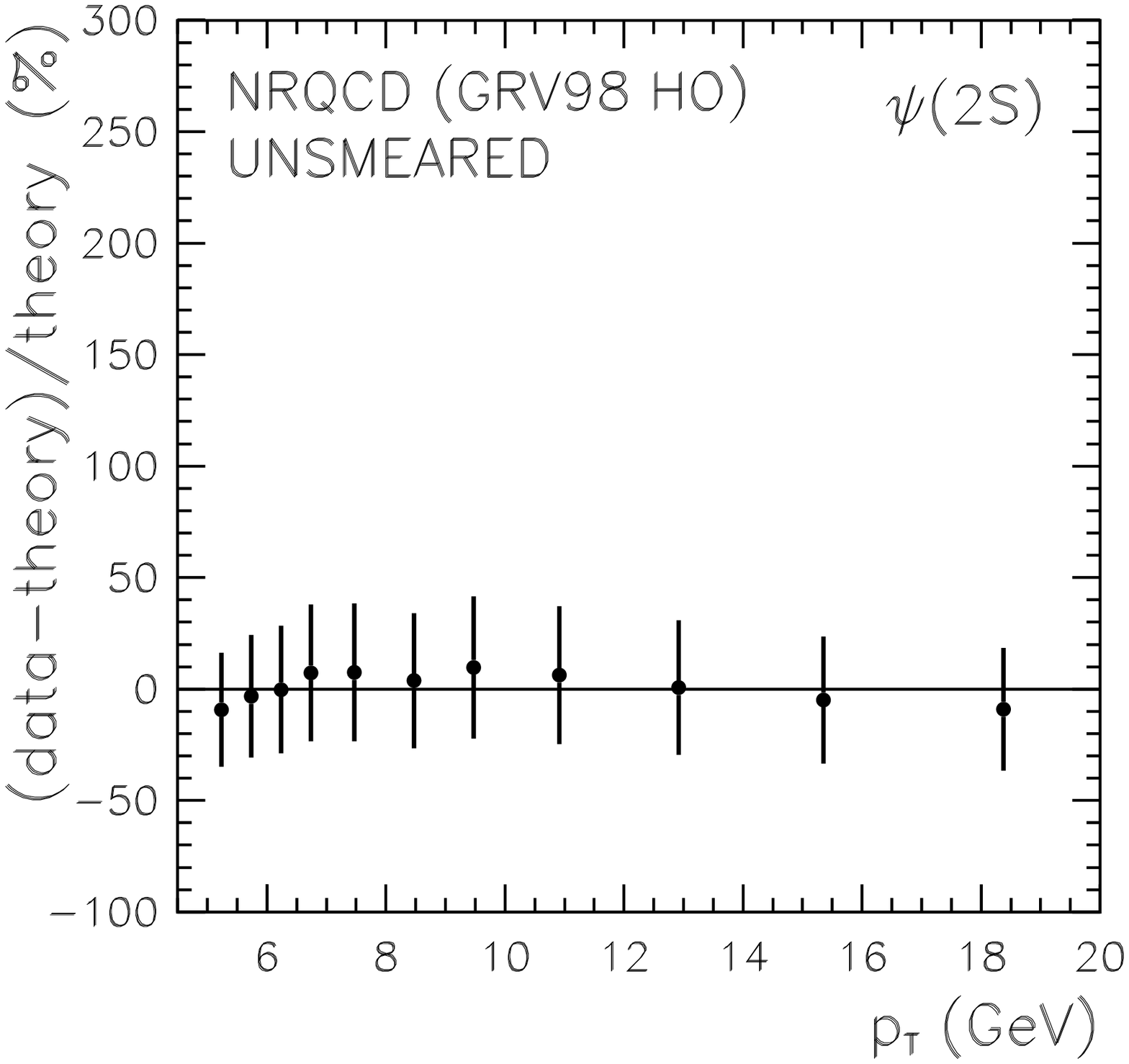}  
\end{tabular}
\caption{$\psi(2S)$ production: $({\rm data}-{\rm theory})/{\rm theory}$.
The plots are as in Fig.~\ref{fig:psi-unsmeared}, but for $\psi(2S)$.}
\label{fig:psip-unsmeared}
\end{figure}

Plots of $({\rm data}-{\rm theory})/{\rm theory}$ for the CEM and NRQCD
factorization predictions without $k_T$ smearing for $\psi(2S)$
production are shown in Fig.~\ref{fig:psip-unsmeared}. Owing to their
larger error bars, the $\psi(2S)$ data have less discriminating power
than the $J/\psi$ data. Nevertheless, it can be seen that the CEM
predictions for $\psi(2S)$ production fit the data poorly, while the
NRQCD factorization predictions fit the data almost too well. (It should
be remembered that the error bars on the CDF data reflect both
systematic and statistical uncertainties.) As in the case of $J/\psi$
production, the CEM predictions have overall normalizations that are too
low and slopes that are too positive. Again, the discrepancies in the
slopes are expected from the fact that the CEM relation (\ref{r-S-wave})
over-estimates the size of $\langle{\cal
O}_8^{\psi(2S)}({}^3S_1)\rangle$ relative to $M_r^{\psi(2S)}$. Even if
the normalizations of the CEM predictions were adjusted to fit the data,
the fits would not be satisfactory.

\begin{figure}
\begin{tabular}{cc}
\includegraphics[width=8cm]{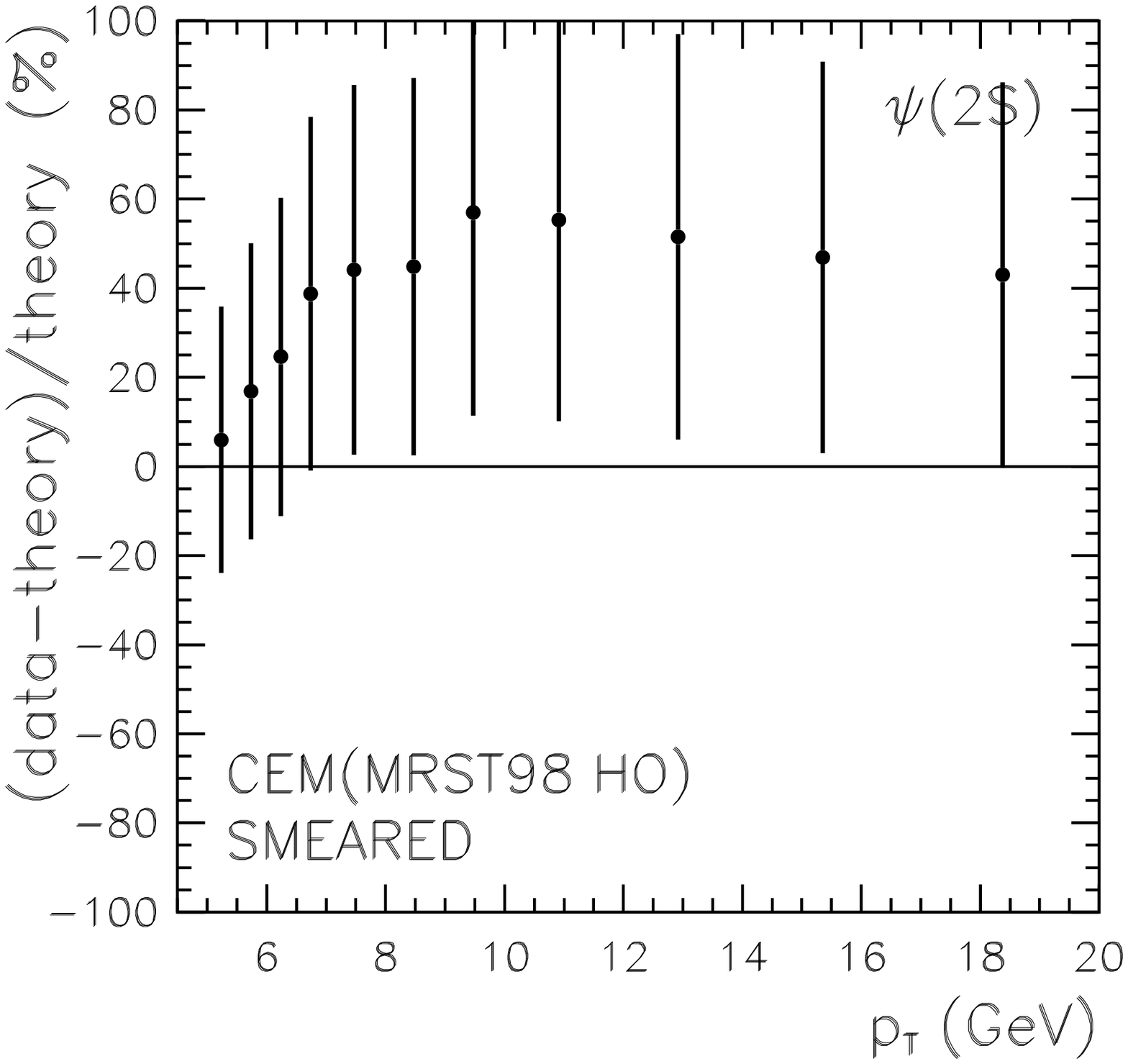}& 
\includegraphics[width=8cm]{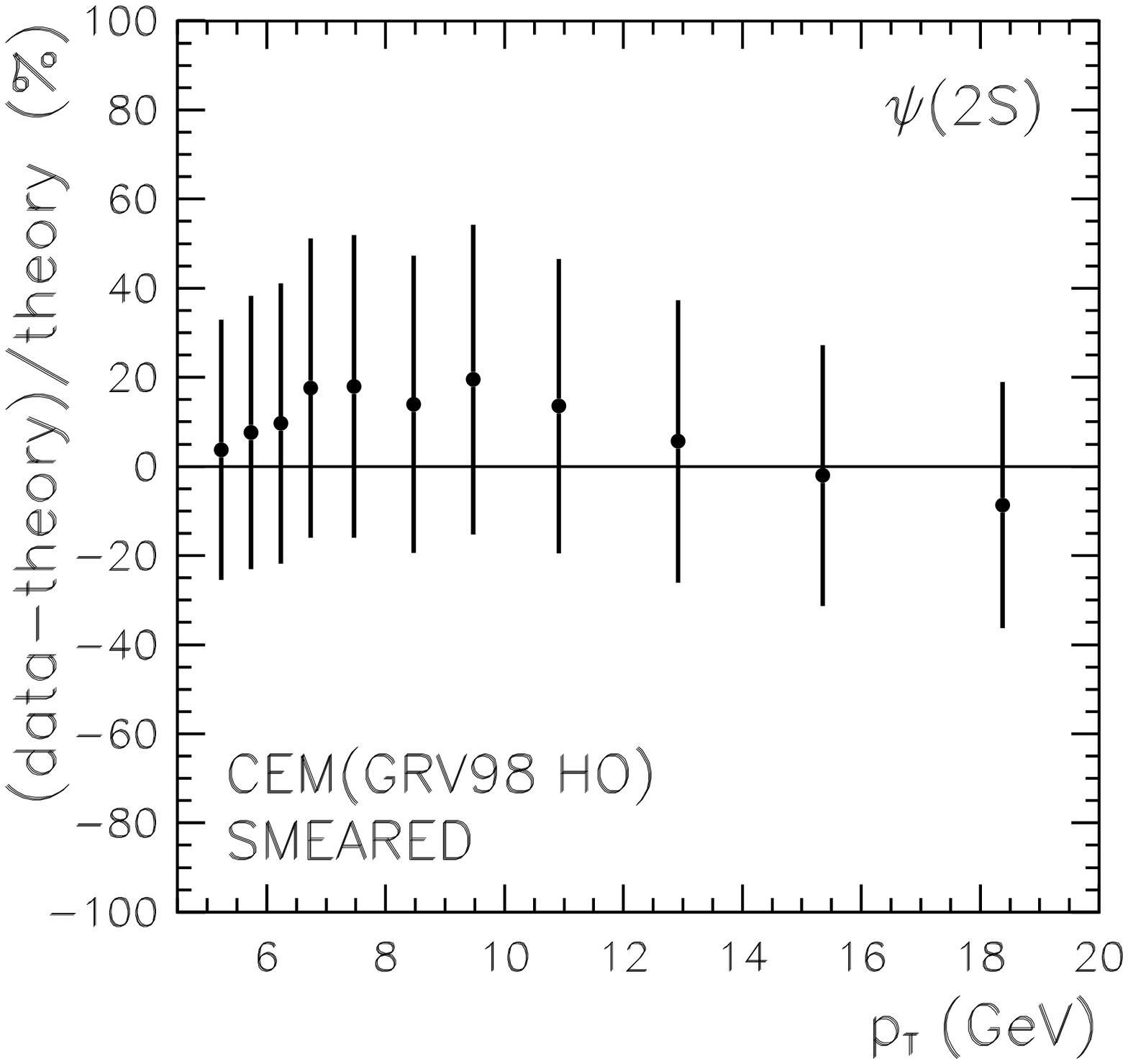}\\
\includegraphics[width=8cm]{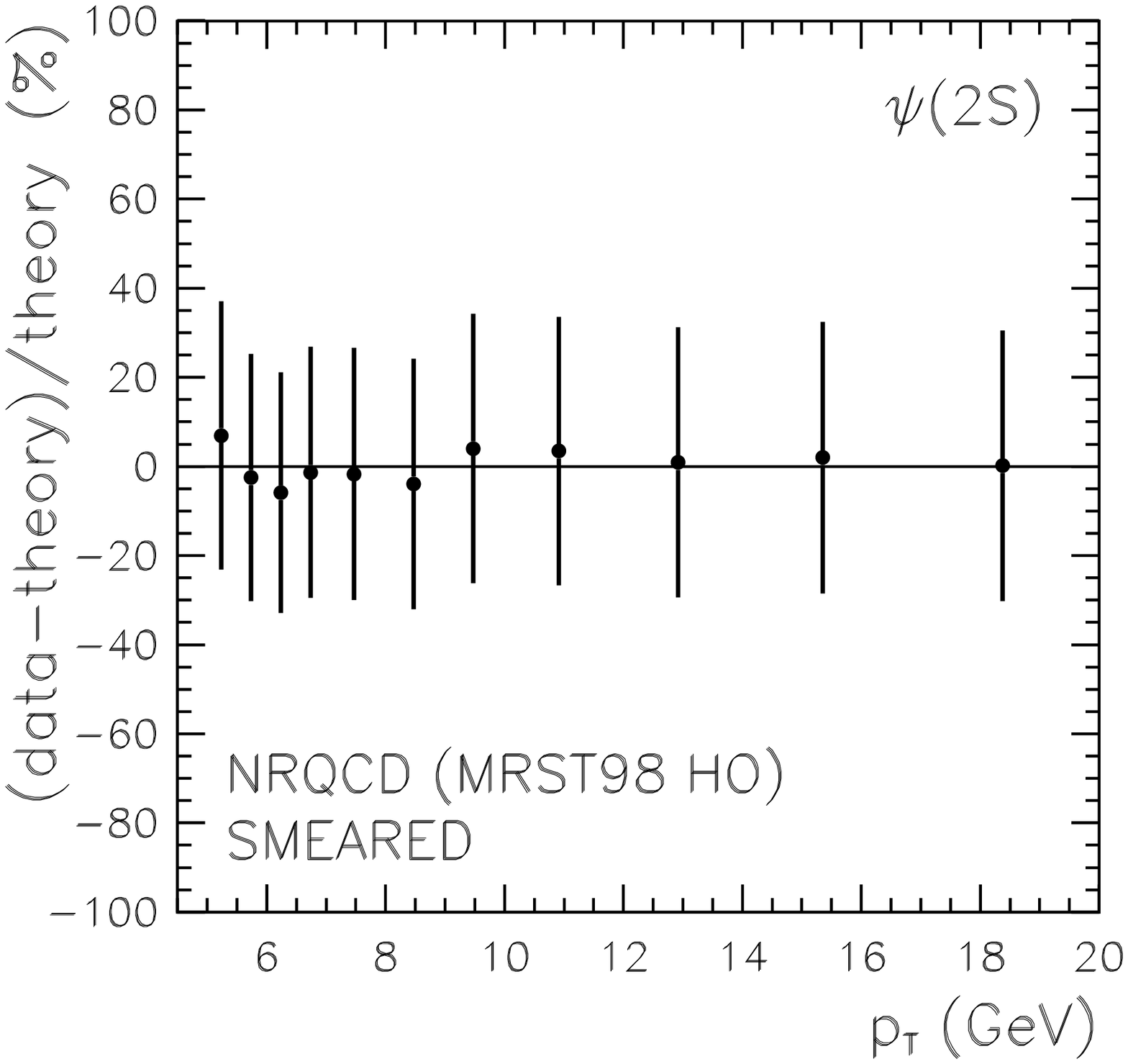}& 
\includegraphics[width=8cm]{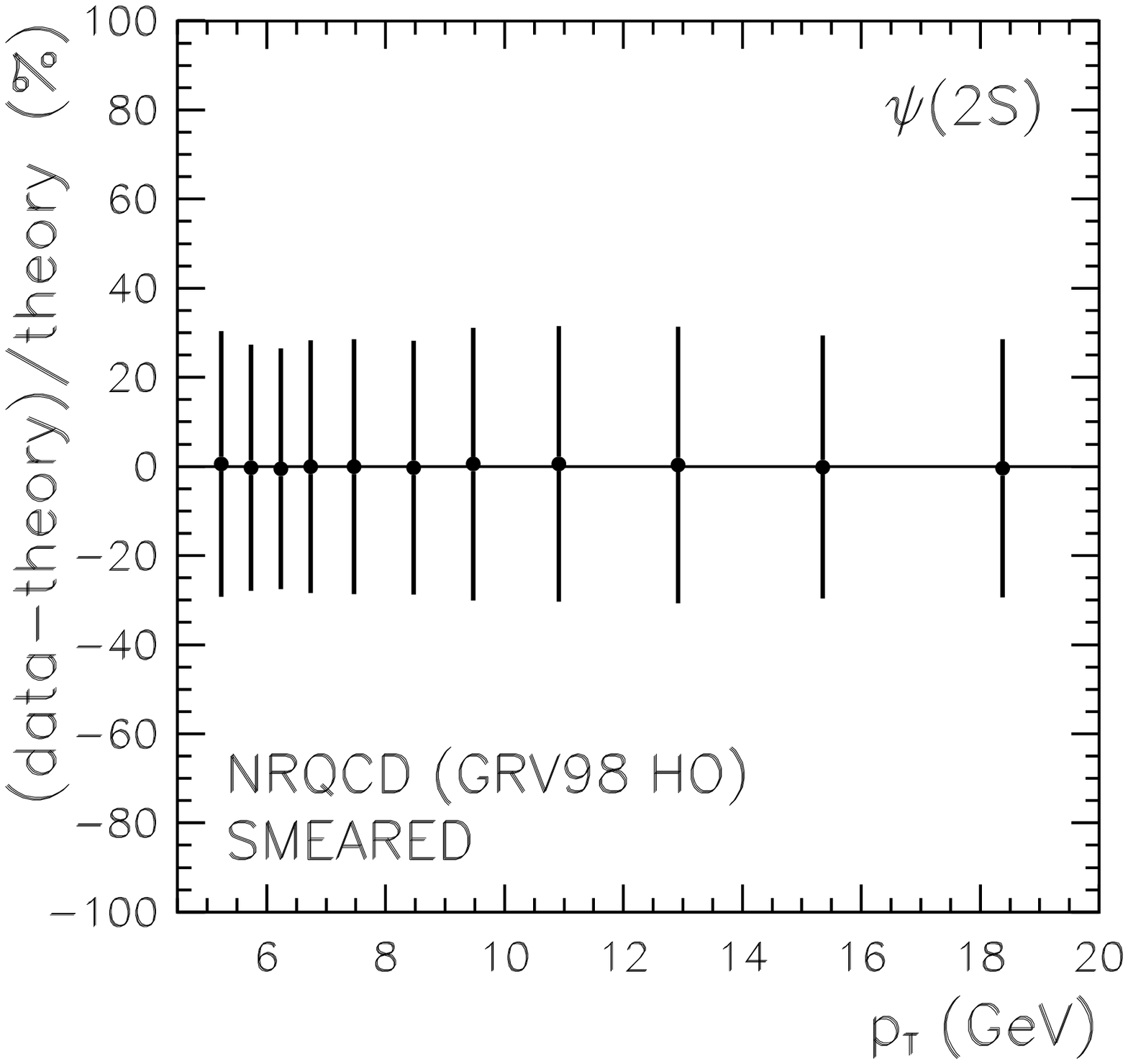}  
\end{tabular}
\caption{$\psi(2S)$ production: $({\rm data}-{\rm theory})/{\rm 
theory}$, with $k_T$ smearing of the theory, as described in the text. 
The plots are as in Fig.~\ref{fig:psi-smeared}, but for $\psi(2S)$.}
\label{fig:psip-smeared}
\end{figure}

Plots of $({\rm data}-{\rm theory})/{\rm theory}$ 
for the $k_T$-smeared CEM and NRQCD
factorization predictions for $\psi(2S)$ production are shown in
Fig.~\ref{fig:psip-smeared}. The effect of $k_T$ smearing is to improve 
both the CEM and NRQCD factorization fits. The CEM fits are improved in 
both normalization and slope and are now compatible with the data.
The matrix elements and $\chi^2/{\rm d.o.f.}$ from the NRQCD
factorization fits and CEM fits to the $\psi(2S)$ data are given in
Table~\ref{tab:psip-fits}. The values of $R^{\psi(2S)}$ in
Table~~\ref{tab:psip-fits} are much greater than $R_{\rm
CEM}^{\psi(2S)}\approx 0.46$ in the fits without $k_T$ smearing and
consistent with $R_{\rm CEM}^{\psi(2S)}$ (and with zero) in the fits
with $k_T$ smearing. The relative values of $R^{\psi(2S)}$ and $R_{\rm
CEM}^{\psi(2S)}$ are consistent with the large discrepancy between the
slopes of the NRQCD factorization and CEM fits without $k_T$ smearing
and the approximate agreement of the slopes of the NRQCD factorization
and CEM fits with $k_T$ smearing.
\begin{table}
\caption{Values of matrix elements, $R^{\psi(2S)}$, and $\chi^2/{\rm
d.o.f.}$ from the NRQCD factorization and CEM fits to the $\psi(2S)$
data. In the NRQCD factorization fits, we set $\langle{\cal
O}^{\psi(2S)}_1({}^3S_1)\rangle$=0.76~GeV$^3$ and give the fitted values
of $\langle{\cal O}^{\psi(2S)}_8({}^3S_1)\rangle$ and
$M_{3.5}^{\psi(2S)}$.}
\label{tab:psip-fits}
\begin{ruledtabular}
\begin{tabular}{l|cccr}
PDF &$\langle{\cal O}^{\psi(2S)}_8({}^3S_1)\rangle$
&$M_{3.5}^{\psi(2S)}$
&$R^{\psi(2S)}$
&$\chi^2/\textrm{d.o.f.}$   \\
&(GeV${}^3\times 10^{-3})$&(GeV${}^3\times 10^{-4})$\\
\hline\hline
\multicolumn{5}{c}{NRQCD Factorization}\\
\hline
MRST98~HO          &2.34  $\pm$ 0.47
                   &44.0  $\pm$ 19.2
                   &\hspace{-1.8ex}18.83 $\pm$ 9.08
                   & 0.35/(11$-$2)=0.04\\
GRV98~HO           & 2.51  $\pm$ 0.52
                   & 55.4  $\pm$ 22.2
                   & \hspace{-1.8ex}22.02 $\pm$ 9.93
                   &0.55/(11$-$2)=0.06\\
MRST98~HO (smeared) & 2.12  $\pm$ 0.26
                   & \hspace{-1.8ex}$-$6.77 $\pm$ 2.20
                   & \hspace{-2.5ex}$-$3.19 $\pm$ 1.11
                   &0.17/(11$-$2)=0.02\\
GRV98~HO (smeared)  &2.34 $\pm$ 0.29
                   & \hspace{-1.8ex}$-$6.80 $\pm$ 2.39
                   & \hspace{-2.5ex}$-$2.90 $\pm$ 1.08
                   &0.22/(11$-$2)=0.02\\
\hline\hline
\multicolumn{5}{c}{Color-Evaporation Model}\\
\hline 
MRST98~HO           &&&& 47.72/11=4.34\\
GRV98~HO            &&&& 29.85/11=2.71\\
MRST98~HO (smeared) &&&& 10.43/11=0.95\\
GRV98~HO (smeared)  &&&& 1.49/11=0.14
\end{tabular}
\end{ruledtabular}
\end{table}

\section{Analysis of $\bm{P}$-wave Charmonium Production}
\label{sec:p-wave}

Now let us turn to the case of production of the $P$-wave charmonium
states $\chi_{cj}$ ($j=0,1,2$) at the Tevatron at transverse momenta
$p_T > 5$ GeV. It is known phenomenologically that the most important
NRQCD matrix elements
are the color-singlet matrix elements $\langle{\cal
O}_1^{\chi_{cj}}({}^3P_j)\rangle$ and the color-octet matrix elements
$\langle{\cal O}_8^{\chi_{cj}}({}^3S_1)\rangle$. The three
color-singlet matrix elements can be expressed in terms of $\langle{\cal
O}_1^{\chi_{c0}}({}^3P_0)\rangle$, and the three color-octet matrix
elements can be expressed in terms of $\langle{\cal
O}_8^{\chi_{c0}}({}^3S_1)\rangle$ by making use of the heavy-quark
spin-symmetry relations
\begin{equation}
\langle{\cal O}_{1,8}^{\chi_{cj}}({}^3P_j)\rangle=(2j+1)\langle{\cal
O}_{1,8}^{\chi_{c0}}({}^3P_0)\rangle,
\end{equation}
which hold up to corrections of order $v^2$. Therefore, we define a 
ratio
\begin{equation}
R^{\chi_c}=
\frac{\langle{\cal O}_8^{\chi_{c0}}({}^3S_1)\rangle}
     {\langle{\cal O}_1^{\chi_{c0}}({}^3P_0)\rangle/m^2}.
\label{eq:R-chi}
\end{equation}
The relation (\ref{me-cem}) yields the CEM prediction
\begin{equation}
R_{\rm CEM}^{\chi_c}=
15C_F \frac{m^2}{k_{\rm max}^2}.
\label{eq:R-chi-CEM}
\end{equation}
The velocity-scaling rules of NRQCD predict that the ratio $R^{\chi_c}$
in Eq.~(\ref{eq:R-chi}) scales as $v^0$. In contrast, we see that the
CEM prediction in Eq.~(\ref{eq:R-chi-CEM}) scales as $v^{-2}$.
Furthermore, with the standard normalization of the NRQCD matrix
elements in Ref.~\cite{Bodwin:1994jh}, the color factor in $R^{\chi_c}$
is estimated \cite{Petrelli:1997ge} to be $1/(2N_c)=1/6$, while the CEM
prediction is that the color factor in $R^{\chi_Q}$ is
$C_F=4/3$.
Both the
discrepancy in the velocity scaling and the discrepancy in the color
factor have the effect of increasing the size of the CEM prediction
relative to the expectation from NRQCD.

In Table~\ref{tab:chi-c} we show values of $R^{\chi_c}$ that were
obtained from several different sets of NRQCD matrix elements that have
been extracted from the transverse-momentum distribution of $\chi_c$'s
produced at the Tevatron. Again, the matrix elements were taken from the
compilation of Ref.~\cite{Kramer:2001hh}. 
\begin{table}
\caption{Values of $R^{\chi_c}$, as defined in Eq.~(\ref{eq:R-chi}), in
the NRQCD factorization approach and in the CEM. As in
Table~\ref{tab:J/psi}, except for $\chi_c$.
The column labeled ``$R_{\rm CEM}^{\chi_c}$''
gives the CEM ratios from Eq.~(\ref{eq:R-chi-CEM})
for the values of  $m_c$  that were used in the NRQCD
extractions of $R^{\chi_c}$.
}
\label{tab:chi-c}
\begin{ruledtabular}
\begin{tabular}{c|cc|ccc}
Reference                & 
\multicolumn{2}{c|}{PDF} & 
$R^{\chi_{c}}$     &
$R^{\chi_{c}}_{\rm CEM}$      &
$m_c$~(GeV)\\
\hline\hline
\multicolumn{6}{c}{\mbox{LO collinear factorization}} \\ 
\hline 
\cite{Cho:1996ce} & \multicolumn{2}{c|}{MRS(D0)~\cite{Martin:1993zi}} & 
 (6.6 $\pm$ 0.8)$\times 10^{-2}$ &  36 & 1.48 \\
\hline \cite{Kniehl:1999qy}& 
\multicolumn{2}{c|}{\mbox{CTEQ4L~\cite{Lai:1997mg}}} & 
 (0.71 $\pm$ 0.21)$\times 10^{-2}$ & 40 & 1.55 \\
\hline                   & 
\multicolumn{2}{c|}{\mbox{MRST-LO(98)~\cite{Martin:1998sq}}} & 
 (5.8 $\pm$ 1.1)$\times 10^{-2}$ &  & \\
\raisebox{2ex}[-2ex]{\cite{Braaten:2000qk}} & 
\multicolumn{2}{c|}{\mbox{CTEQ5L~\cite{Lai:1999wy}}} & 
 (4.7 $\pm$ 0.8)$\times 10^{-2}$
 &  \raisebox{2ex}[-2ex]{37} &\raisebox{2ex}[-2ex]{1.5}\\
\end{tabular}
\end{ruledtabular}
\end{table}
The values of $R^{\chi_c}$ in Table~\ref{tab:chi-c} are reasonably close
to the value $R^{\chi_c}\approx v^0/(2N_c)\approx 0.17$ that one would
expect on the basis of the velocity-scaling rules and the estimate of
the color factor. Table~\ref{tab:chi-c} also contains the CEM
values $R_{\rm CEM}^{\chi_c}$. As expected, they are much larger
than the values of $R^{\chi_c}$ that follow from the 
data.\footnote{Because the
CEM prediction scales incorrectly with $v$, we expect the disagreement
between the CEM  ratio and the value extracted from  the data to be 
even more dramatic  for $R^{\chi_b}$ than for $R^{\chi_c}$. 
For $m_b=4.7$~GeV and $m_B=5.28$~GeV, 
the CEM  ratio predicted by  Eq.~(\ref{eq:R-chi-CEM}) is 
 $R_{\rm CEM}^{\chi_b} \approx 76$, 
while the expectation from NRQCD is that 
 $R^{\chi_b}$ should be comparable to $R^{\chi_c}$.} Consequently,
we expect the CEM to predict a cross section that is relatively too 
large at high $p_T$, where the contribution from $\langle{\cal
O}_8^{\chi_{c0}}({}^3S_1)\rangle$ dominates, in comparison with the
cross section at low $p_T$, where the contribution from $\langle{\cal
O}_1^{\chi_{c0}}({}^3P_0)\rangle$ dominates.

A compilation of values  of matrix
elements, values of $R^{\chi_c}$, and $\chi^2/{\rm d.o.f.}$ from the
NRQCD factorization and CEM fits to the $\chi_c$ data are given in
Table~\ref{tab:chi-fits}. The values of $R^{\chi_c}$ in
Table~~\ref{tab:chi-fits} are much less than $R_{\rm CEM}^{\chi_c}
\approx 37$.

\begin{figure}
\begin{tabular}{cc}
\includegraphics[width=7.5cm]{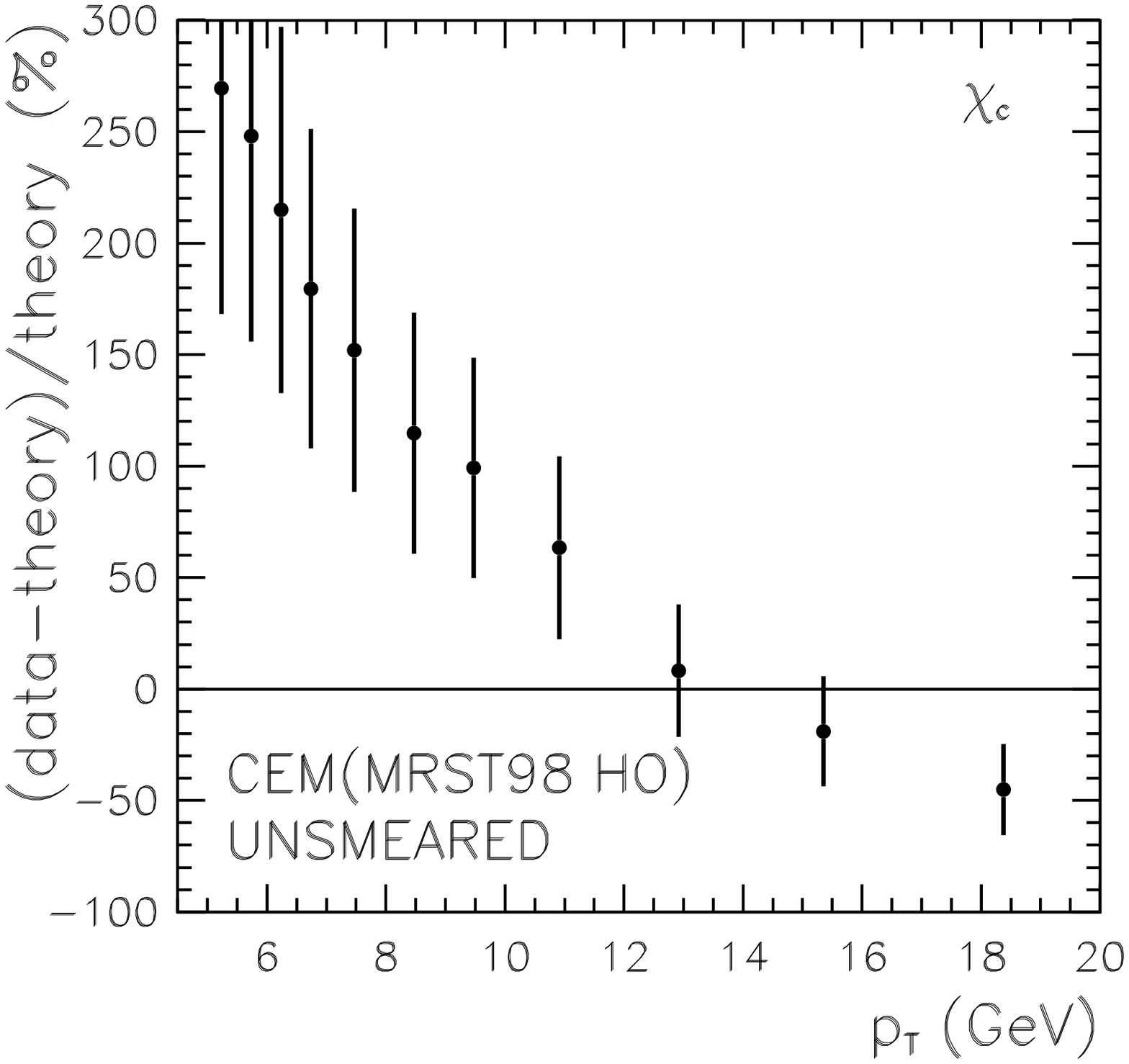}&
\includegraphics[width=7.5cm]{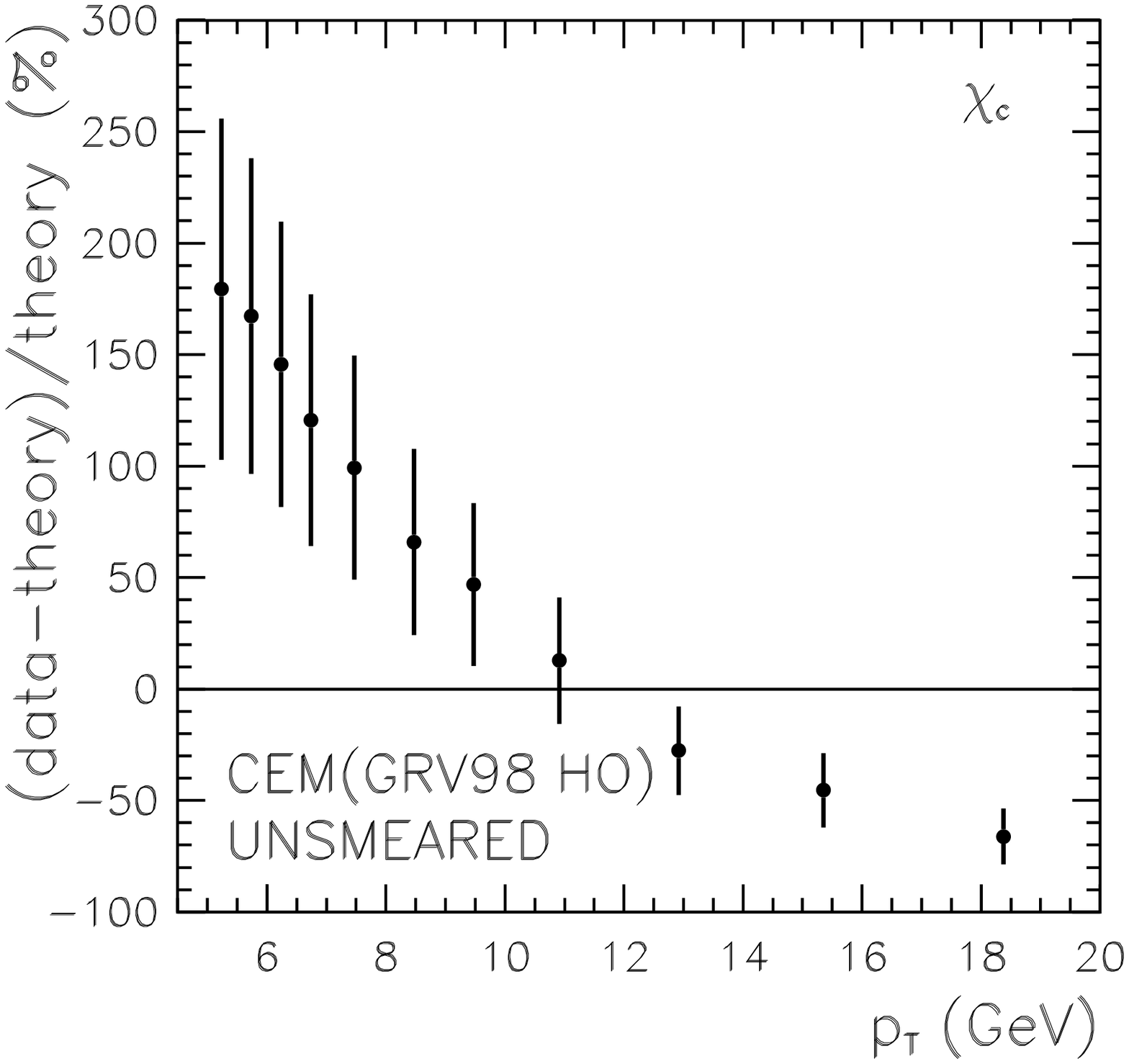} 
\vspace{-1cm}\\
\includegraphics[width=7.5cm]{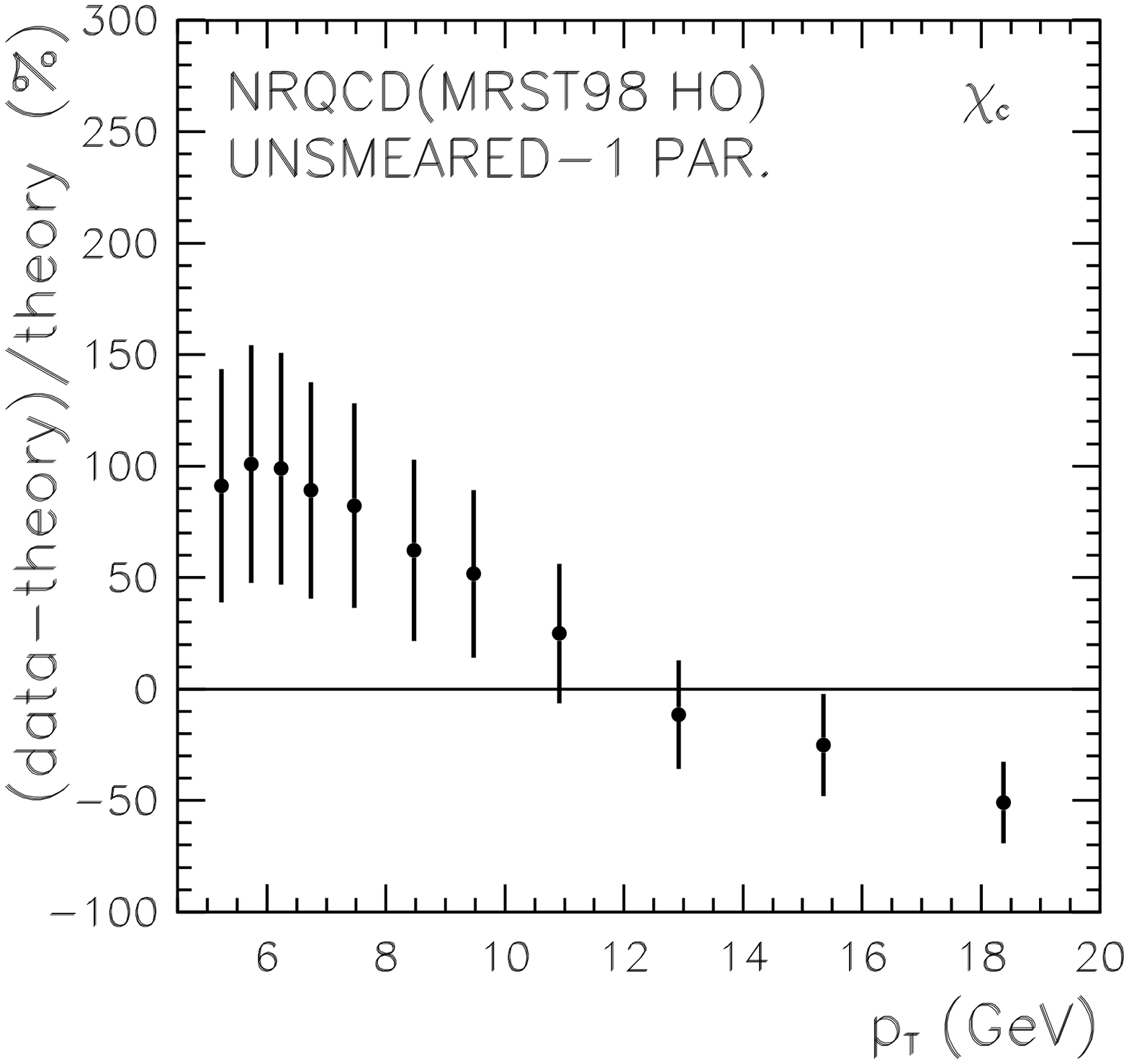}&
\includegraphics[width=7.5cm]{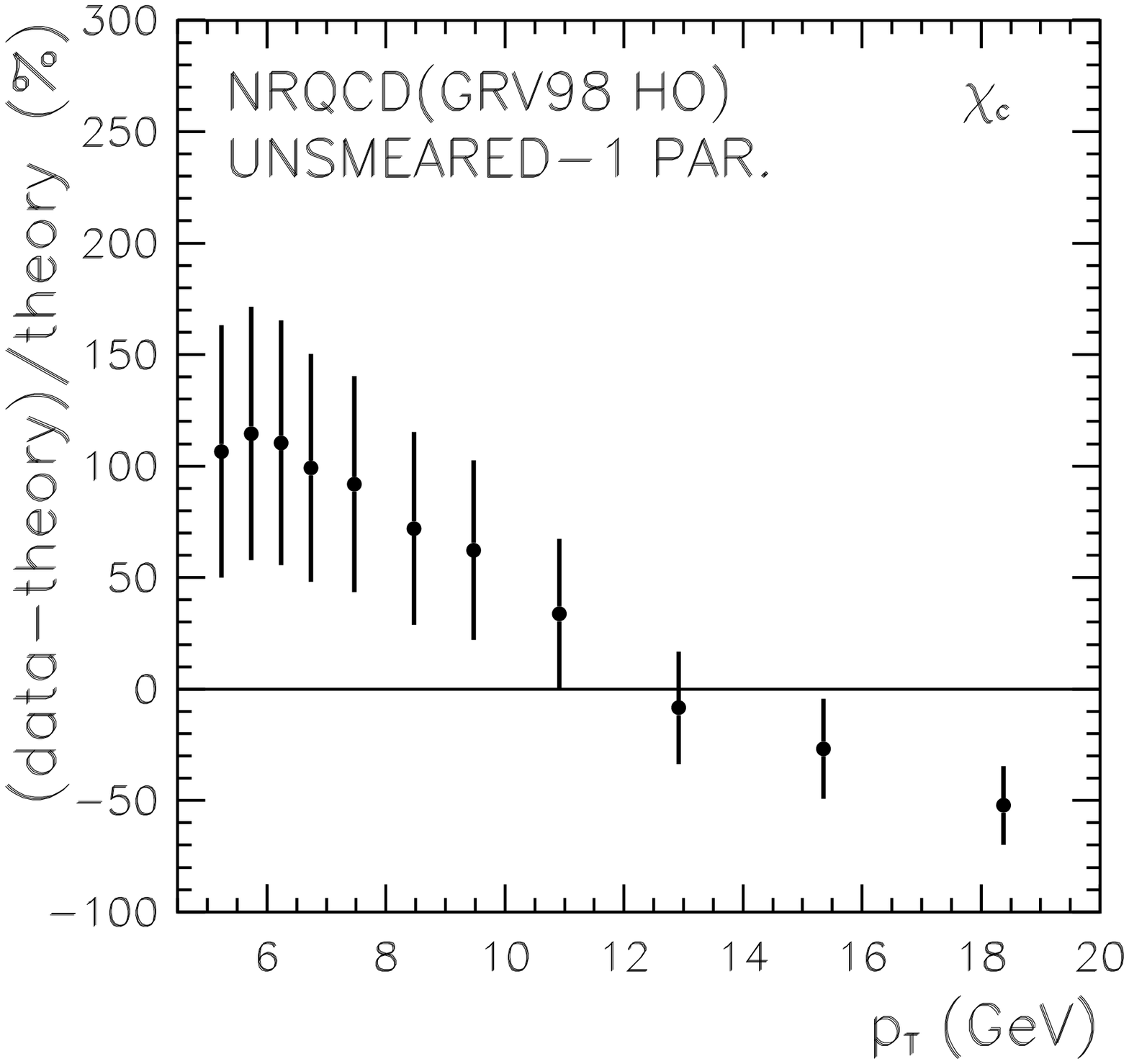} 
\vspace{-1cm}\\
\includegraphics[width=7.5cm]{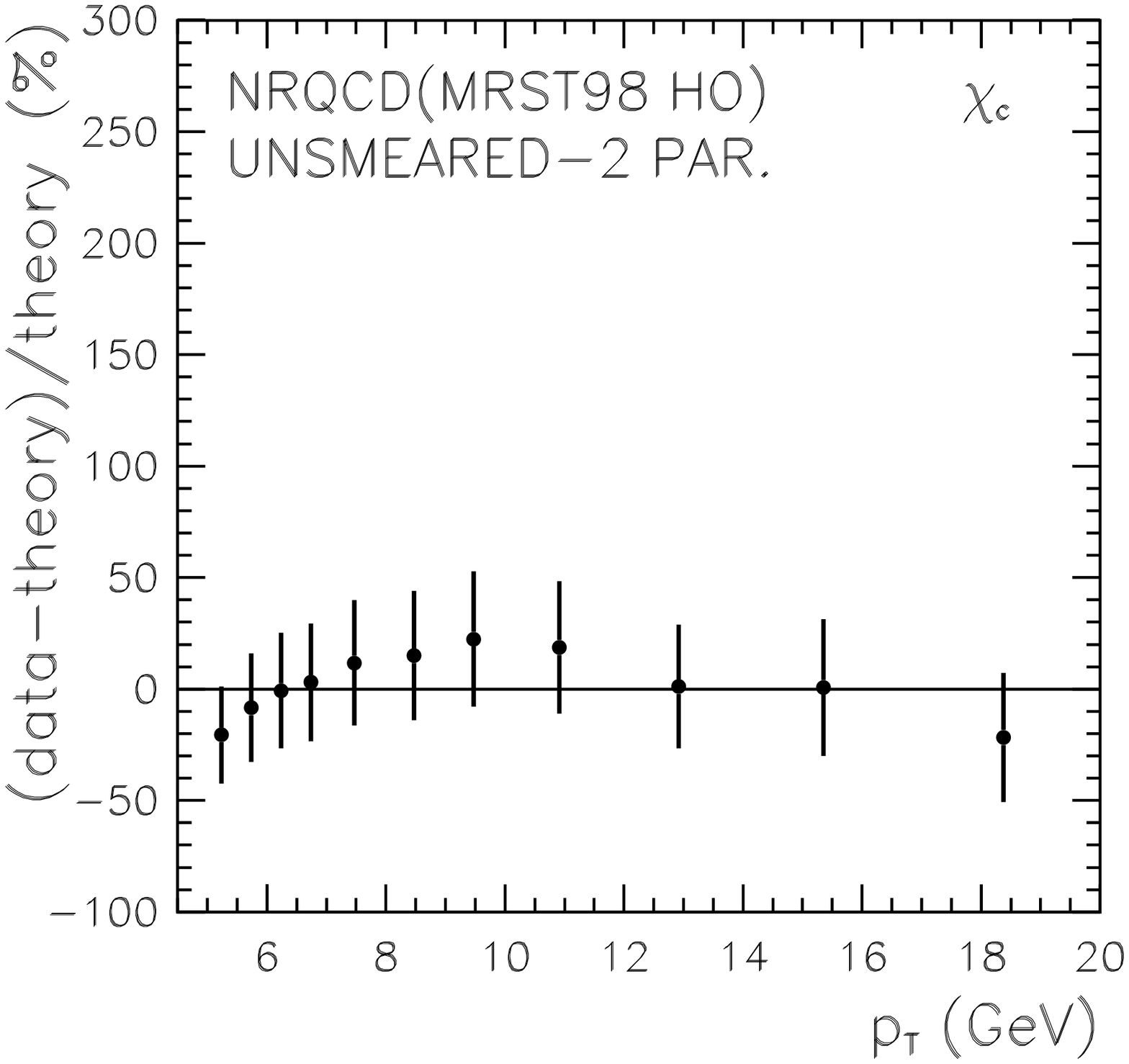}&
\includegraphics[width=7.5cm]{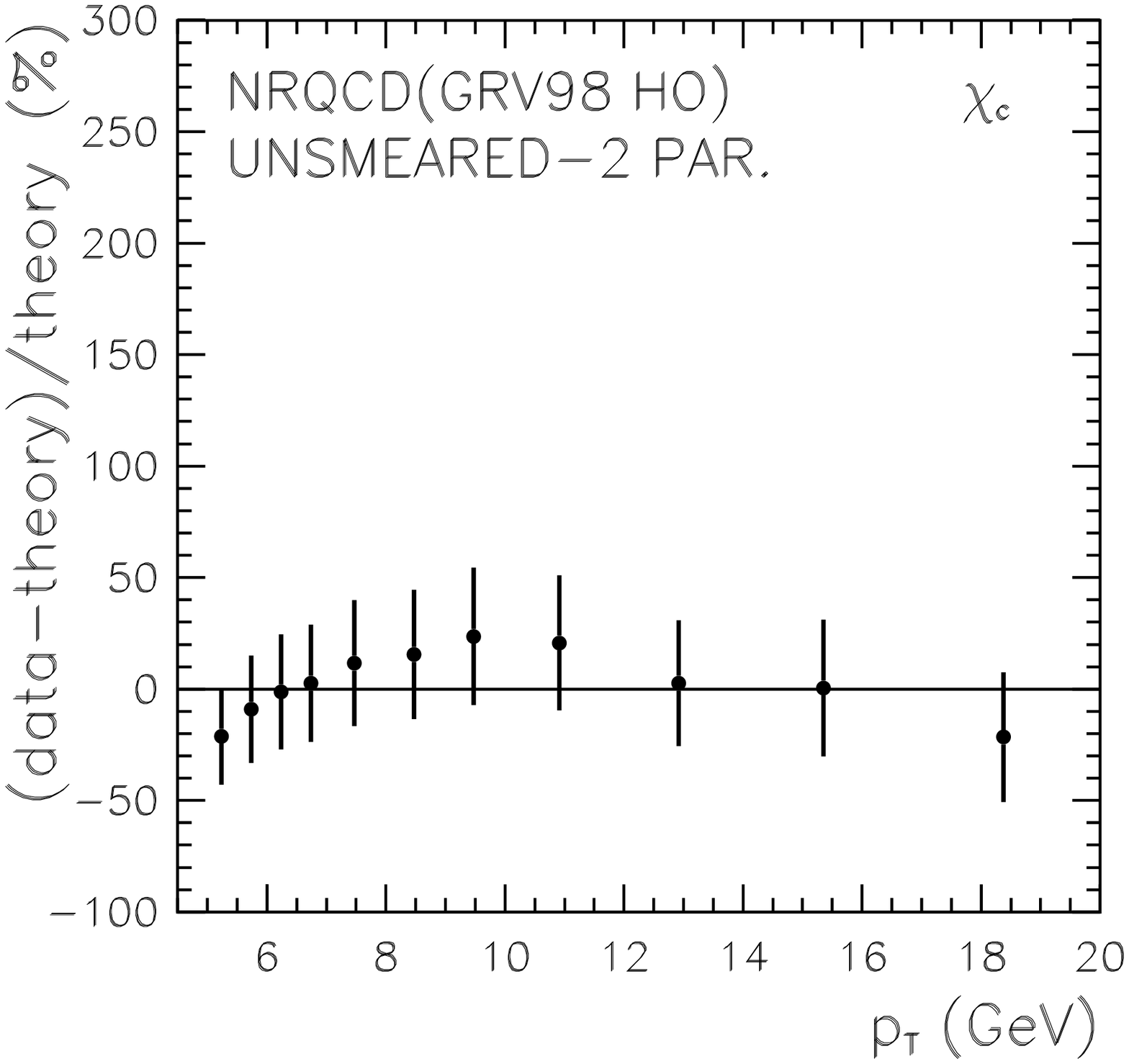} 
\end{tabular}
\caption{$\chi_c$ production: $({\rm data}-{\rm theory})/{\rm theory}$.
The plots are as in Fig.~\ref{fig:psi-unsmeared}, except that they are for
$\chi_c$, the top row is for the CEM predictions, and the middle
and bottom rows are for the 1-parameter and 2-parameter NRQCD
factorization predictions, respectively.}
\label{fig:chi-unsmeared}
\end{figure}

Plots of $({\rm data}-{\rm theory})/{\rm theory}$ for the CEM and NRQCD
factorization predictions without $k_T$ smearing for $\chi_c$ production
are shown in Fig.~\ref{fig:chi-unsmeared}.
In the NRQCD factorization fits in the middle row, we have fixed the
value of $\langle{\cal O}^{\chi_{c0}}_1({}^3P_0)\rangle$ to be
$7.2\times 10^{-2}$~GeV$^5$. This value is taken from a global fit to
the existing data for inclusive decays of $\chi_c$ states
\cite{Maltoni:2000km}, which determines the corresponding decay matrix
element to be $\langle\chi_{c0}|{\cal O}_1({}^3P_0)|\chi_{c0}\rangle=
(7.2\pm 0.9)\times 10^{-2}$~GeV$^5$. (Color-singlet decay and production
matrix elements are simply related, up to corrections of order $v^4$
\cite{Bodwin:1994jh}.) In the NRQCD factorization fits in the lower row,
we have treated $\langle{\cal O}^{\chi_{c0}}_1({}^3P_0)\rangle$ as a
free parameter. From the fits, it can be seen that both the CEM and
1-parameter NRQCD factorization predictions have normalizations that are
too small and slopes that are too positive relative to the data. Both
sets of fits show substantial disagreements with the slope of the data.
However, the discrepancies are considerably greater in the CEM fits than
in the NRQCD factorization fits. The large differences in slope between
the NRQCD factorization and CEM predictions are consistent with the
values of $R^{\chi_c}$ in Table~~\ref{tab:chi-fits} relative to $R_{\rm
CEM}^{\chi_c}$. The 2-parameter NRQCD factorization predictions fit the
data much better than the 1-parameter predictions, but the values of
$\langle{\cal O}^{\chi_{c0}}_1({}^3P_0)\rangle$ that are obtained are
about a factor of six larger than the phenomenological value of the
corresponding decay matrix element. Even given the large theoretical
uncertainties in the determination of the decay matrix element, this
discrepancy is unacceptably large. Thus, we conclude that the
disagreement of the unsmeared NRQCD factorization prediction with the
$\chi_c$ data cannot be ameliorated by treating $\langle{\cal
O}^{\chi_{c0}}_1({}^3P_0)\rangle$ as a free parameter.

\begin{figure}
\begin{tabular}{cc}
\includegraphics[width=7.5cm]{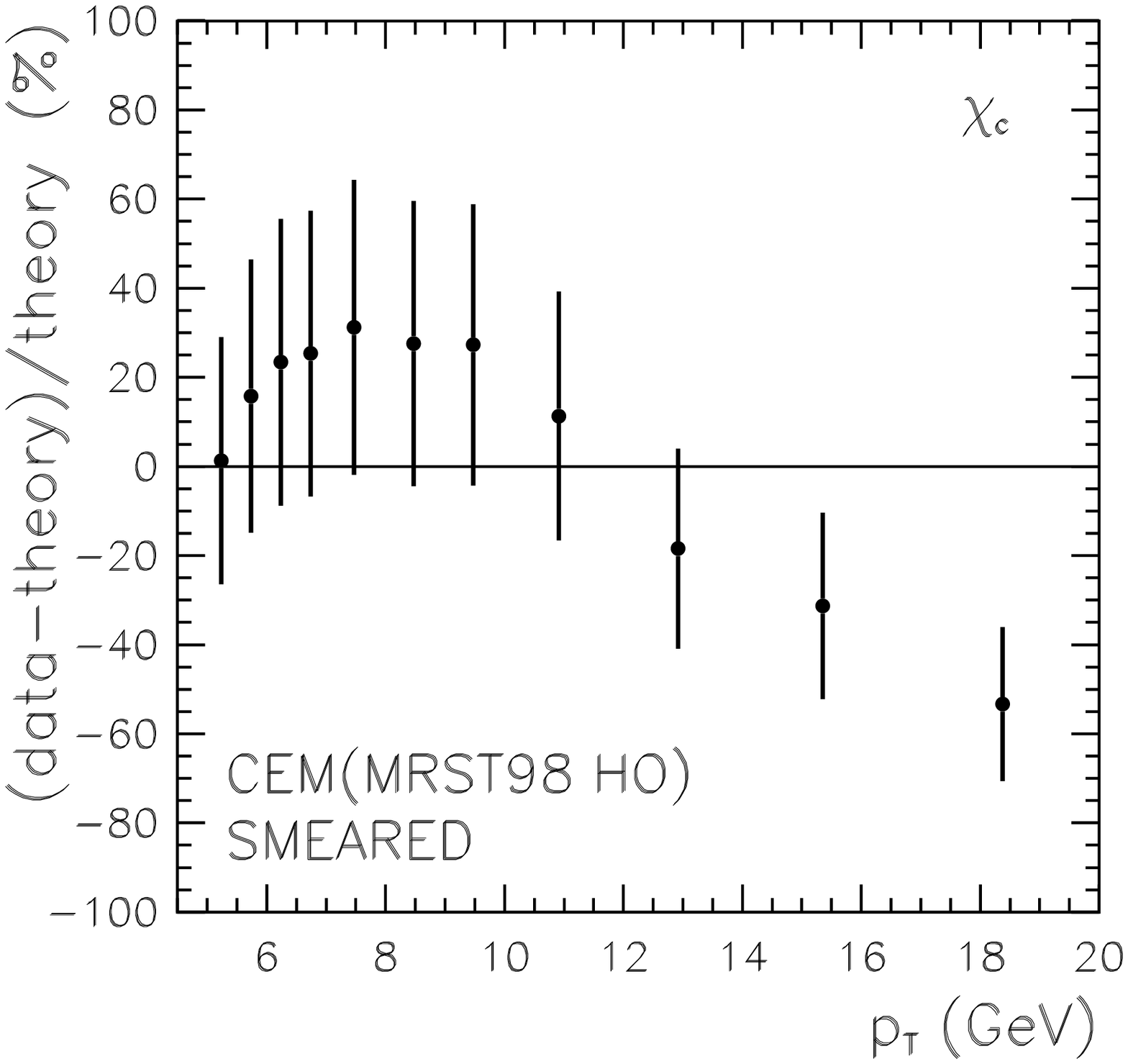}&
\includegraphics[width=7.5cm]{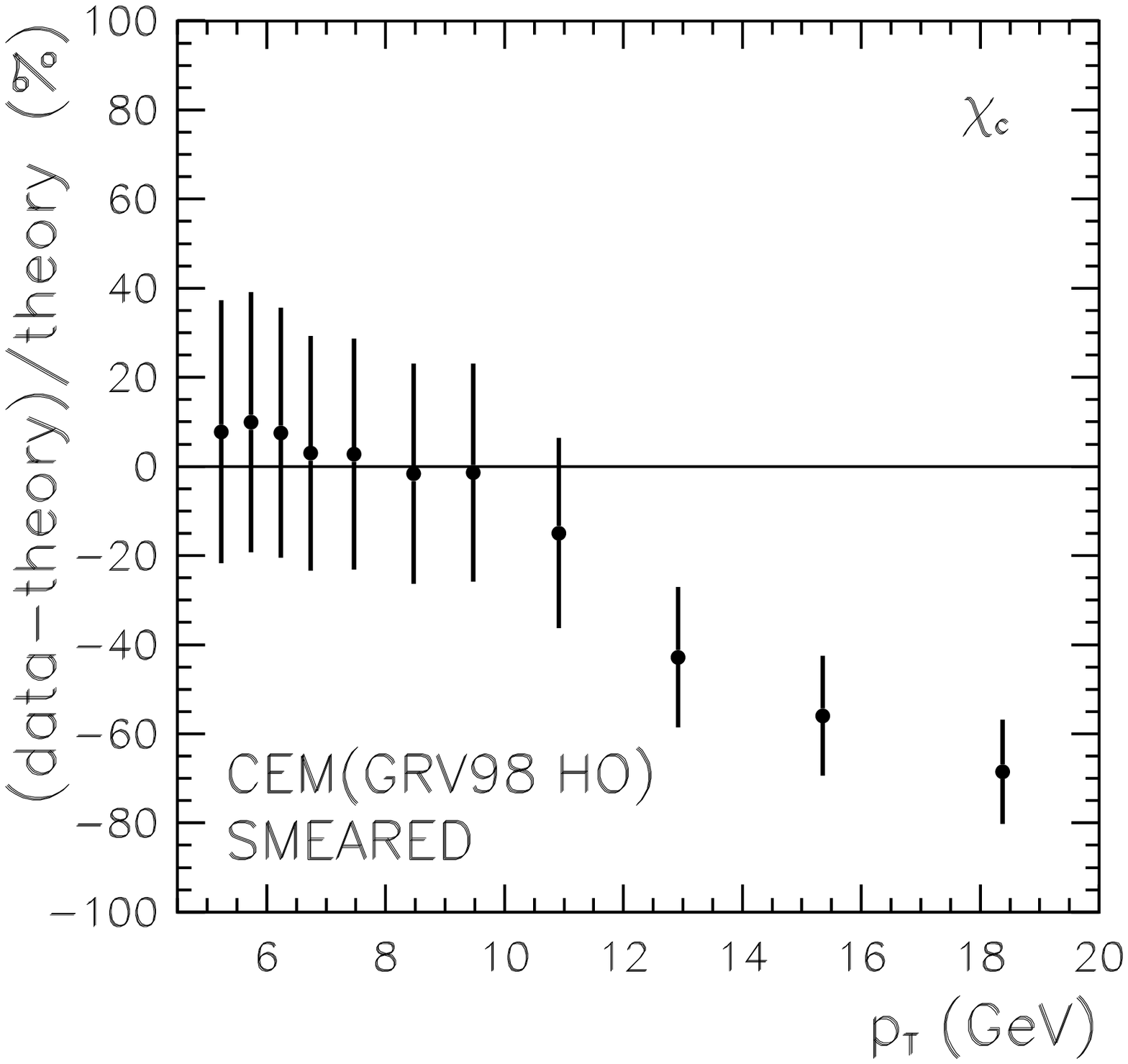} 
\vspace{-1cm}\\                          
\includegraphics[width=7.5cm]{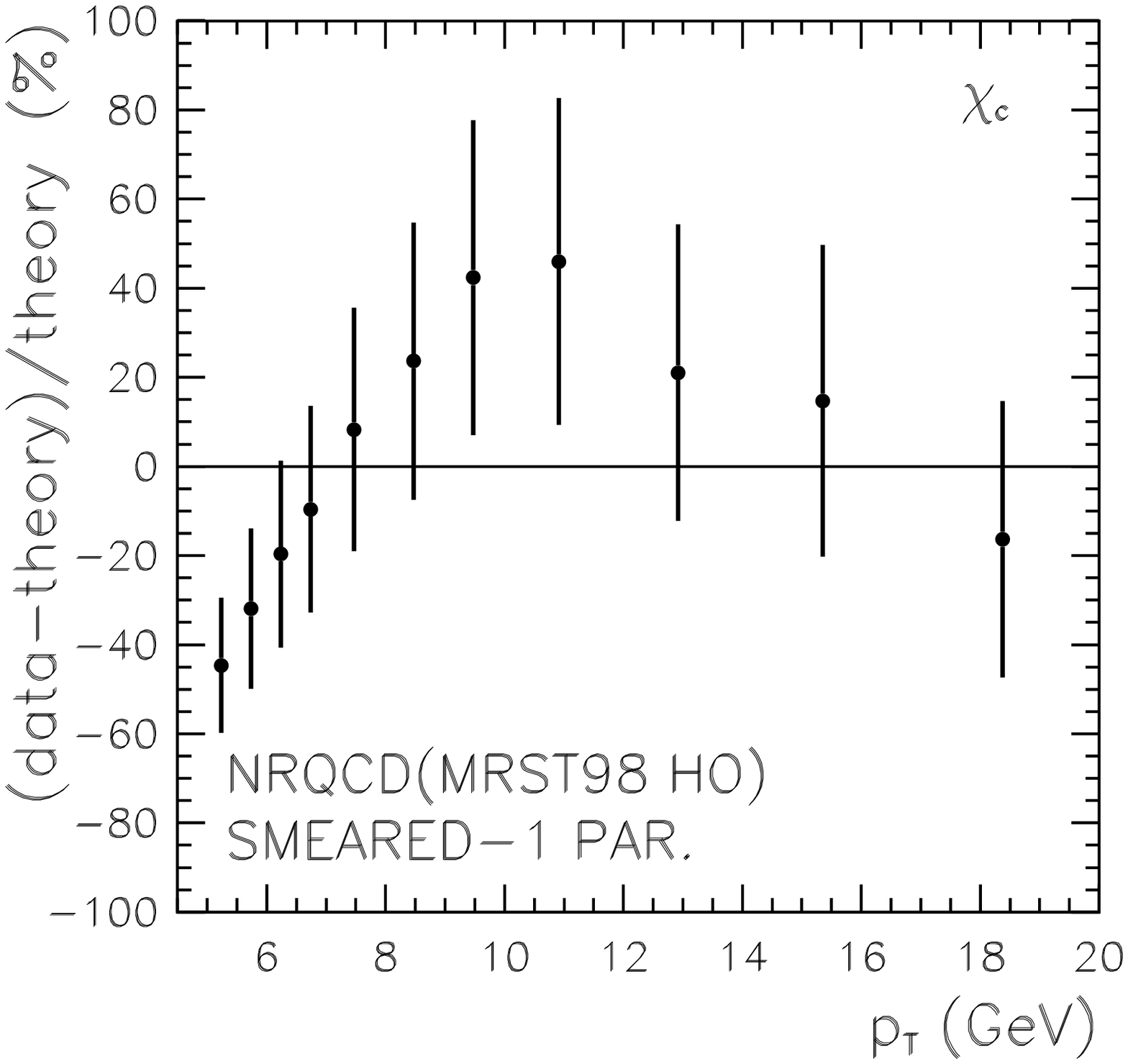}&
\includegraphics[width=7.5cm]{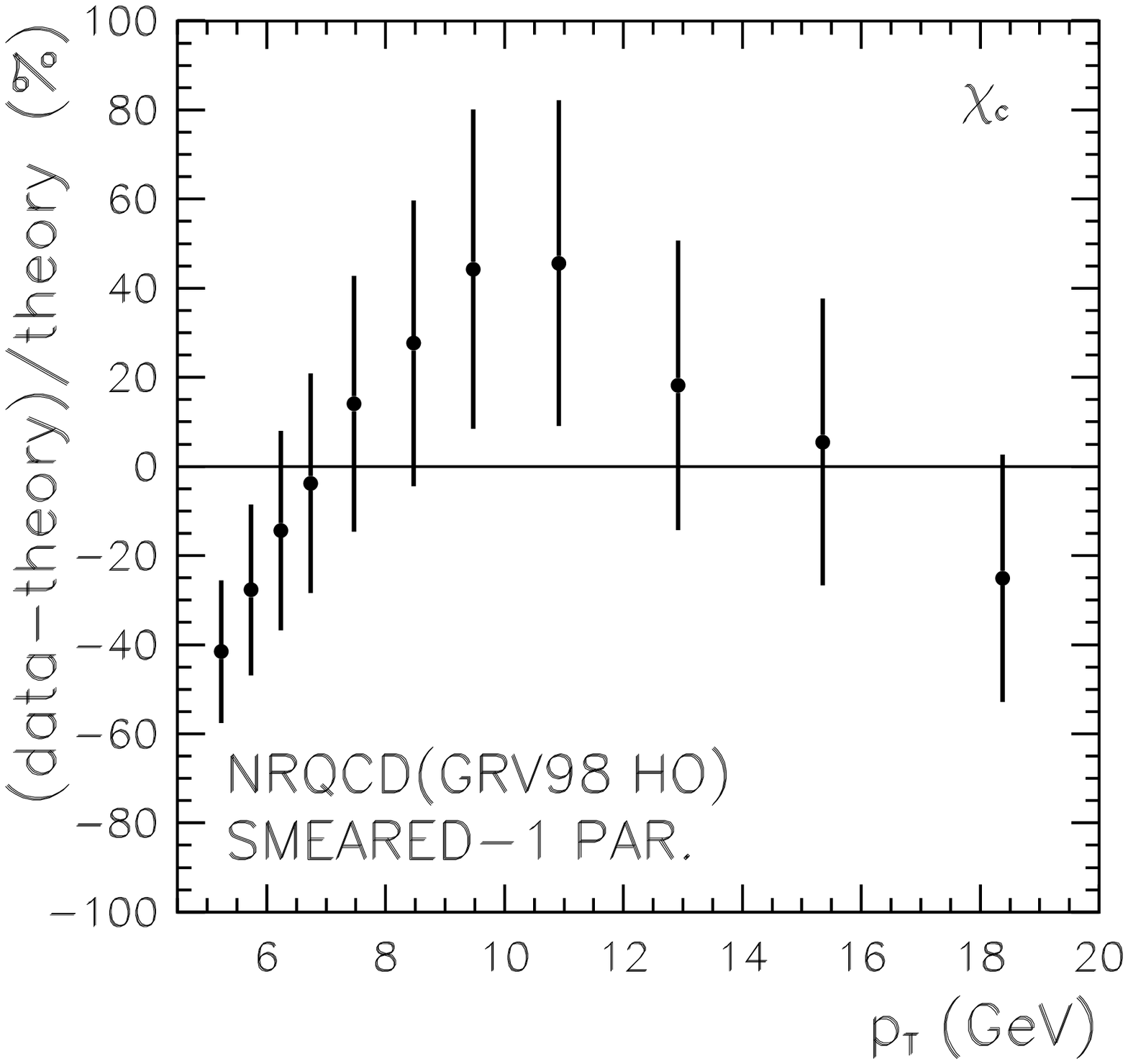} 
\vspace{-1cm}\\                          
\includegraphics[width=7.5cm]{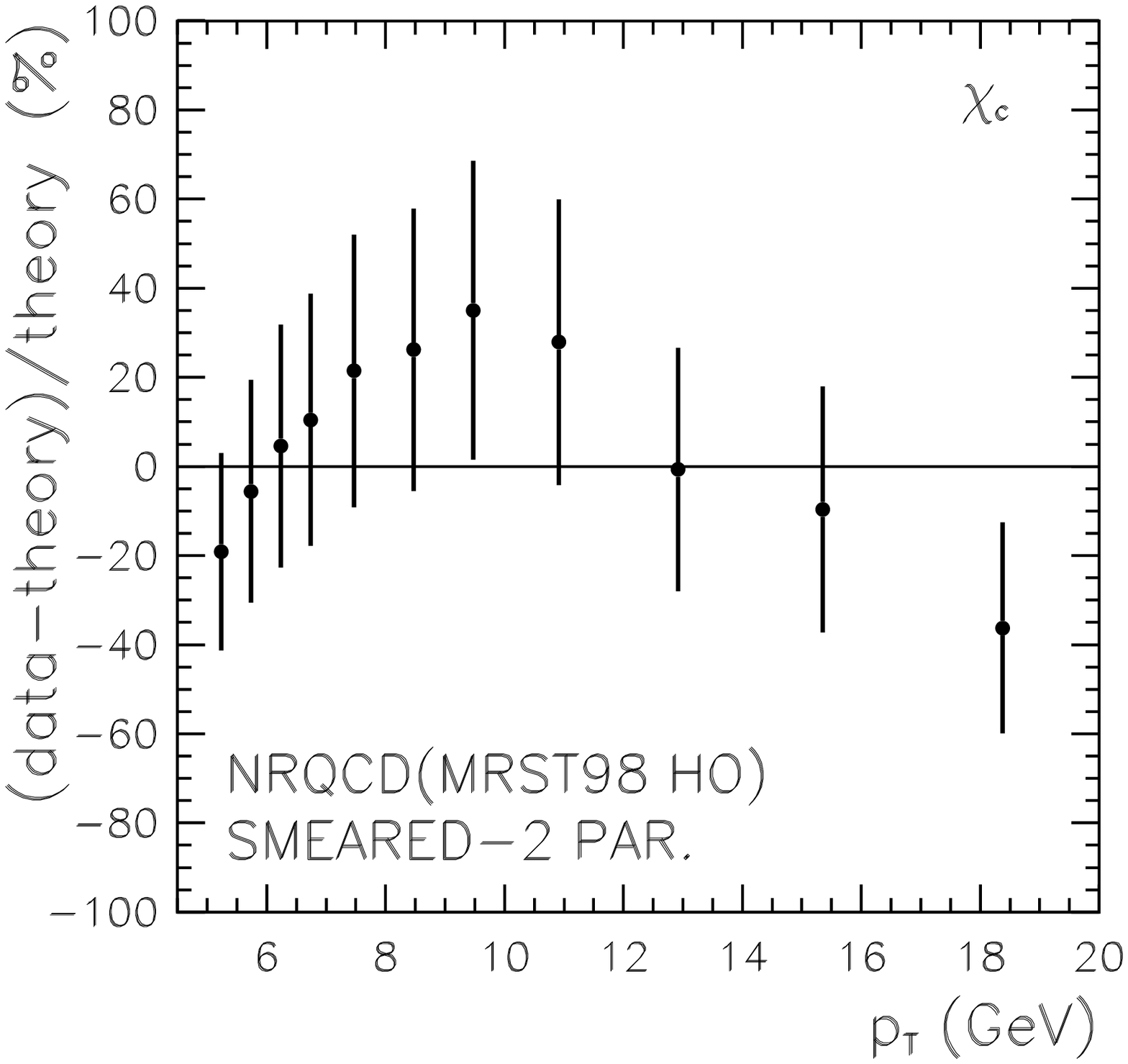}&
\includegraphics[width=7.5cm]{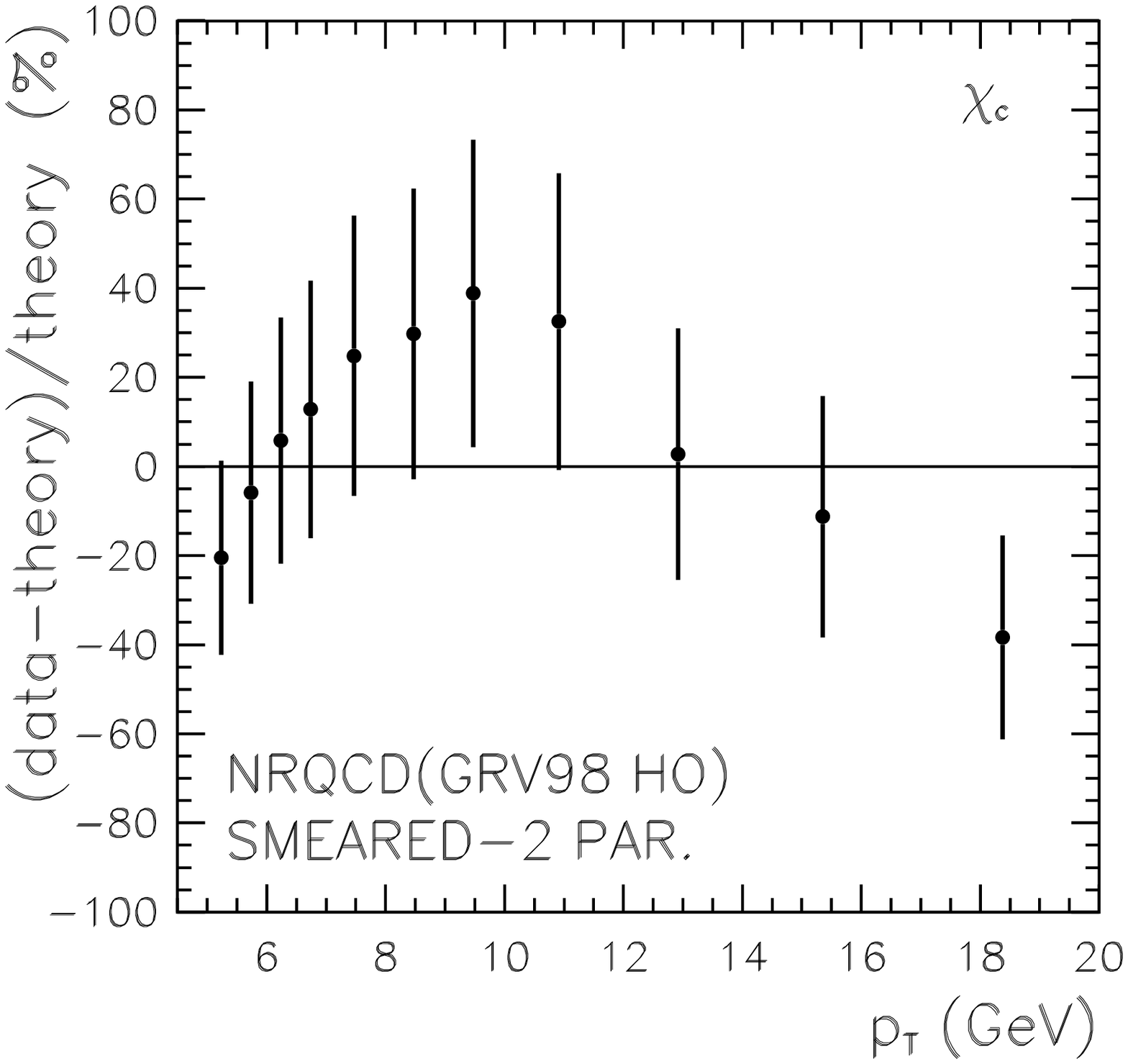} 
\end{tabular}
\caption{$\chi_c$ production: $({\rm data}-{\rm theory})/{\rm theory}$,
with $k_T$ smearing of the theory, as described in the text. The plots
are as in Fig.~\ref{fig:psi-smeared}, except that they are for $\chi_c$,
the top row is for the CEM predictions, and the middle and bottom
rows are for the 1-parameter and 2-parameter NRQCD factorization
predictions, respectively.}
\label{fig:chi-smeared}
\end{figure}

Plots of $({\rm data}-{\rm theory})/{\rm theory}$ for the $k_T$-smeared 
CEM and NRQCD factorization predictions for $\chi_c$ production are shown in
Fig.~\ref{fig:chi-smeared}.
$k_T$ smearing improves the quality of the 1-parameter NRQCD
factorization fits and the quality of the CEM fit that is based on the
MRST98~HO parton distributions, but does not substantially change the
overall quality of the 2-parameter NRQCD factorization fits or the CEM
fit that is based on the GRV98~HO parton distributions. The CEM
predictions are still too large at high $p_T$, and the 1-parameter NRQCD
factorization predictions are now too large at low $p_T$. The
2-parameter NRQCD factorization predictions with $k_T$ smearing give
good fits to the data, but the fitted values of $\langle{\cal
O}^{\chi_{c0}}_1({}^3P_0)\rangle$ now are considerably {\it smaller} than the
central value from $\chi_c$ decays. The relative values of $R^{\chi_c}$
and $R_{\rm CEM}^{\chi_c}$ are consistent with the relative slopes of
the CEM and NRQCD predictions. The improvement in the 1-parameter NRQCD 
factorization fits with $k_T$ smearing suggests that multiple gluon 
emission may play an important role in quarkonium production. However, 
the remaining discrepancy between the NRQCD
factorization 1-parameter fits and the data and the small value of
$\langle{\cal O}^{\chi_{c0}}_1({}^3P_0)\rangle$ in the NRQCD
factorization 2-parameter fits suggest that the value $\langle
k_T^2\rangle=2.5~\hbox{GeV}^2$ is too large to be compatible with NRQCD
factorization. In the fits to the $J/\psi$ and $\psi(2S)$ production
data, one could compensate to some extent for changes in the value of
$\langle k_T^2\rangle$ by adjusting the value of $M_{3.5}^H$. However,
the corresponding adjustment of $\langle{\cal
O}^{\chi_{c0}}_1({}^3P_0)\rangle$ in the fits to the $\chi_c$ production
data is significantly constrained by the $\chi_c$ decay data. Hence,
comparisons of the predictions of NRQCD factorization with the $\chi_c$
data may provide a more stringent test of NRQCD factorization than
comparisons with the $J/\psi$ and $\psi(2S)$ data. A definitive test
would require one to replace the $k_T$-smearing model for multiple gluon
emission with a first-principles calculation. 
\begin{table}
\caption{Values of matrix elements, $R^{\chi_c}$, and $\chi^2/{\rm
d.o.f.}$ from the NRQCD factorization and CEM fits to the $\chi_c$ data.
In the NRQCD factorization fits, the upper sets of parameters are for
fits in which $\langle{\cal O}^{\chi_{c0}}_1({}^3P_0)\rangle$ is fixed,
as described in the text, while the lower sets of parameters are for
fits in which $\langle{\cal O}^{\chi_{c0}}_1({}^3P_0)\rangle$ is
varied.}
\label{tab:chi-fits}
\begin{ruledtabular}
\begin{tabular}{l|ccrr}
PDF & $\langle{\cal O}^{\chi_{c0}}_1({}^3P_0)\rangle$
&$\langle{\cal O}^{\chi_{c0}}_8({}^3S_1)\rangle$
&$R^{\chi_c}$
&$\chi^2/\textrm{d.o.f.}$\\
&(GeV${}^5\times 10^{-2}$)&(GeV${}^3\times 10^{-3}$)&($10^{-2}$)&\\
\hline\hline
\multicolumn{5}{c}{NRQCD Factorization}\\
\hline
MRST98~HO          & 7.2~(input) 
                   & 3.59 $\pm$ 0.39 
                   & 11.23 $\pm$ 1.23
                   & 31.0/(11$-$1)=3.10\\
GRV98~HO           & 7.2~(input) 
                   & 3.94 $\pm$ 0.43 
                   & 12.30 $\pm$ 1.35
                   & 35.5/(11$-$1)=3.55\\
MRST98~HO (smeared) & 7.2~(input) 
                   & 1.71 $\pm$ 0.29 
                   & 5.36 $\pm$ 0.89
                   & 17.4/(11$-$1)=1.74\\
GRV98~HO (smeared)  & 7.2~(input) 
                   & 2.08 $\pm$ 0.32
                   & 6.50 $\pm$ 0.99
                   & 14.5/(11$-$1)=1.45\\
\hline
MRST98~HO          & 40.8 $\pm$ 6.3
                   & 1.20 $\pm$ 0.60 
                   & 0.66 $\pm$ 0.35
                   & 2.97/(11$-$2)=0.33\\
GRV98~HO           & 48.7 $\pm$ 7.3 
                   & 1.17 $\pm$ 0.65
                   & 0.54 $\pm$ 0.31
                   & 3.19/(11$-$2)=0.35\\
MRST98~HO (smeared) &\hspace{1.0ex}3.88 $\pm$ 1.00 
                   & 2.43 $\pm$ 0.36 
                   & \hspace{-1.0ex}14.12 $\pm$ 4.21
                   & 6.40/(11$-$2)=0.71\\
GRV98~HO (smeared)  &\hspace{1.0ex}4.39 $\pm$ 1.09
                   & 2.67 $\pm$ 0.39 
                   & \hspace{-1.0ex}13.66 $\pm$ 3.93
                   & 7.88/(11$-$2)=0.88\\
\hline\hline
\multicolumn{5}{c}{Color-Evaporation Model}\\
\hline
MRST98~HO           &&&& 50.20/11=4.56\\
GRV98~HO            &&&& 66.30/11=6.03\\
MRST98~HO (smeared) &&&& 16.15/11=1.47\\
GRV98~HO (smeared)  &&&& 63.69/11=5.79
\end{tabular}
\end{ruledtabular}
\end{table}

\section{Discussion and Conclusions}
\label{sec:conclusions}

In this paper, we have compared the CEM and NRQCD factorization
approaches to inclusive quarkonium production. As we have mentioned, the
predictions of the CEM are at odds with a number of experimental
observations. These include the different fractions of $J/\psi$'s from
$\chi_{c}$ decays that occur in $B$ decays and in prompt production at
the Tevatron, the nonzero polarization of $J/\psi$'s in $e^+e^-$
annihilation at the $B$ factories, the nonzero polarization of
$\Upsilon(2S)$ and $\Upsilon(3S)$ in a fixed-target experiment, and the
deviation from $3/5$ of the ratio of the prompt-production cross
sections for $\chi_{c1}$ and $\chi_{c2}$ at the Tevatron. Nevertheless,
one might hope that the CEM would still be useful for predicting rates
of inclusive quarkonium production at large $p_T$. While some of the
predictions of the NRQCD factorization approach do not agree well with
the data, for example, in the cases of the polarization of $J/\psi$'s
produced at the Tevatron and $J/\psi$ photoproduction 
at the Hadron Electron Ring Accelerator (HERA)
at Deutsches Elektronen-Synchrotron (DESY), the
experimental and theoretical uncertainties are sufficiently large that
one cannot yet make a definite statement about the validity of NRQCD
factorization. (For a comprehensive review of these issues, see
Ref.~\cite{Brambilla:2004wf}.)

By making use of the NRQCD factorization expressions for the quarkonium
production cross section and for the perturbative $Q\bar Q$ production
cross section, we have translated the CEM assumptions into  predictions
for the ratios of  NRQCD production matrix elements. In some
instances, these predictions are at odds with the velocity-scaling rules
of NRQCD. In other cases, they disagree with ratios of the
nonperturbative NRQCD production matrix elements that have been
extracted from phenomenology. Both of these facts indicate that the CEM
picture for the evolution of a $Q\bar Q$ pair into a quarkonium state is
very different from that of NRQCD.

A comparison of  the CEM ratios with the phenomenological ratios  that
have been extracted from the CDF data indicates that the CEM predicts a
ratio  $M_r^H/\langle{\cal O}_8^{H}({}^3S_1)\rangle$ that is too small in
$J/\psi$ and $\psi(2S)$ production and a ratio $\langle{\cal
O}_8^{\chi_{Q0}}({}^3S_1)\rangle/\langle{\cal
O}_1^{\chi_{Q0}}({}^3P_0)\rangle$ that is too large in $\chi_c$
production. Both of these predictions of the CEM would be expected to
lead to cross sections that have too positive a slope, as a function of
$p_T$, relative to the data. This expectation is borne out by
comparisons of the CEM with the CDF data for $J/\psi$, $\psi(2S)$, and
$\chi_c$ production. The CEM predictions do not yield satisfactory
fits to the $J/\psi$, $\psi(2S)$, or $\chi_c$ data. The NRQCD
factorization predictions yield satisfactory fits to the $J/\psi$ and
$\psi(2S)$ data, but not to the $\chi_c$ data, unless one relaxes the
constraint on $\langle{\cal O}^{\chi_{c0}}_1({}^3P_0)\rangle$ that
follows from its relationship to the corresponding decay matrix element,
which in turn is fixed through a global fit to the inclusive $\chi_c$
decay data \cite{Maltoni:2000km}. The normalizations of the CEM
predictions are fixed through comparisons with the fixed-target data for
charmonium production. 

$k_T$ smearing provides a phenomenological model for the effects of
multiple gluon emission from the initial-state partons in a hard
collision. Its effects are to smooth singularities at $p_T=0$ in
fixed-order calculations, to increase the predicted cross section at
moderately low $p_T$ (away from the singular region), and to increase
the predicted cross section by a smaller amount at high $p_T$. Hence,
the inclusion of $k_T$ smearing would be expected to improve the fits of
the CEM predictions to the charmonium data, which it does. Even with
$k_T$ smearing, the CEM predictions show substantial disagreement with
the data for $J/\psi$ and $\chi_c$ production, but agree with the 
$\psi(2S)$ data, which have larger error bars. The smeared NRQCD
factorization predictions are in good agreement with the
data in the $J/\psi$ and $\psi(2S)$ cases and in reasonably good
agreement in the $\chi_c$ case. 

In general, we find that the nonperturbative NRQCD matrix elements can
be adjusted so as to obtain good fits to the data, even when the slopes
and normalizations of the partonic cross sections are modified quite
strongly through the inclusion of $k_T$ smearing or the use of different
parton distributions. Hence, the comparisons of the predictions of NRQCD
factorization with the Tevatron data alone do not provide demanding tests
of the theory. Comparisons with data from other processes are required.
(See Ref.~\cite{Brambilla:2004wf} for a review of such comparisons.) As
we have mentioned, in the case of $\chi_c$ production, the NRQCD
factorization fits are constrained by the relationship of $\langle{\cal
O}^{\chi_{c0}}_1({}^3P_0)\rangle$ to the corresponding decay matrix
element. Thus, there is less freedom in that case to tune the matrix
elements to obtain a good fit to the data than in the cases of $J/\psi$
and $\psi(2S)$ production. Consequently, $\chi_c$ production may
provide a more stringent test of NRQCD factorization. The disagreement
of the unsmeared NRQCD factorization prediction and the reasonable
agreement of the smeared NRQCD factorization prediction with the shape
of the $\chi_c$ production data suggest that, if the NRQCD factorization
picture is valid, then inclusion of the effects of multiple gluon
emission is essential in obtaining the correct shape of the cross
section.

Overall, the CEM predictions do not provide a satisfactory description
of the data. Since the unsmeared CEM predictions for charmonium
production at the Tevatron are absolute, one might reasonably argue that
they could not be expected to fit the data as well as the NRQCD
factorization predictions, which generally involve two free parameters.
However, even if the normalizations of the CEM predictions are adjusted
so as to optimize the fits, the predictions are still incompatible with
the $J/\psi$ and $\chi_c$ data, owing to incompatibilities between the
predicted and observed slopes. In the case of the $k_T$-smeared
predictions, the amount of $k_T$ smearing has been adjusted so as to
optimize the fits to the $J/\psi$ data. Even so, the CEM predictions do
not describe the data well.

Finally, we should mention that there are several important sources of 
uncertainty in the theoretical predictions. These include uncertainties
in the choices of factorization and renormalization scales,
uncertainties in the parton distributions, uncertainties in the value of
$m_c$, and uncertainties from uncalculated corrections of higher order
in $\alpha_s$. The uncertainty from uncalculated higher-order
corrections is probably the largest. It is often estimated by comparing
the differences between the predictions that are obtained by varying the
factorization and renormalization scales by a factor of two. Such an
exercise has been carried out in the case of the NRQCD factorization
predictions for $J/\psi$ and $\psi(2S)$ production \cite{Beneke:1997yw}
and suggests that the error from uncalculated higher-order corrections
may be as large as 100\% at high $p_T$ and 40\% at low $p_T$. The CEM
predictions presented in this paper involve a simultaneous change in
scale and in parton distributions. If we assume that the change in scale
is the dominant effect, then the uncertainty from it is about 75\% at
high $p_T$ and 10\% at low $p_T$. It is likely that the large
uncertainty in the NRQCD factorization prediction could be accommodated
by a change in the values of the NRQCD matrix elements, except in the
case of the $\chi_c$ predictions. Even if we take the larger
uncertainties in the NRQCD factorization prediction as being indicative
of the uncertainties in the CEM predictions, that would not be enough to
make the unsmeared CEM predictions compatible with the data. Therefore,
it seems likely that, once the effects of multiple gluon emission are
taken into account properly, through calculations of higher-order
corrections rather than through the $k_T$-smearing model, the CEM
predictions will still show a serious incompatibility with the data. As
we have mentioned, calculations of the effects of multiple gluon
emission are crucial to the NRQCD factorization predictions for $\chi_c$
production, as well. Consequently, they could play an important role in
sharpening the comparison between the CEM and NRQCD factorization.

\begin{acknowledgments}
We thank Ramona Vogt for providing numerical tables of the CEM
predictions for charmonium production. We also thank
F.~Maltoni, M.~Mangano, and A.~Petrelli for providing their computer
codes for calculating the NLO quarkonium cross sections and the
evolution of the fragmentation contribution. Work by G.~T.~Bodwin in the
High Energy Physics Division at Argonne National Laboratory is supported
by the U.~S.~Department of Energy, Division of High Energy Physics, under
Contract No.~W-31-109-ENG-38. E.~Braaten is also supported in part by
the Department of Energy under grant DE-FG02-91-ER4069. 
The work of J.~Lee was supported by the KOSEF Basic Research Program under
Grant No. R01-2005-000-10089-0 and by the SK Group under Grant for physics
research at Korea University.
\end{acknowledgments}


\end{document}